\documentstyle[natbib209,2up,multicol,float]{arPrepn}
\twouparticle

\newcommand{\be}{\begin{equation} }
\newcommand{\ee}{\end{equation}}

\newcommand{\bea}{\begin{eqnarray}}
\newcommand{\eea}{\end{eqnarray}}

\floatstyle{plain}
\newfloat{plate}{thp}{pla}
\floatname{plate}{Plate}

\renewcommand{\refname}{Literature Cited}
\def\spose#1{\hbox to 0pt{#1\hss}}
\def\simlt{\mathrel{\spose{\lower 3pt\hbox{$\mathchar"218$}}
     \raise 2.0pt\hbox{$\mathchar"13C$}}}
\def\simgt{\mathrel{\spose{\lower 3pt\hbox{$\mathchar"218$}}
     \raise 2.0pt\hbox{$\mathchar"13E$}}}
\def\simpropto{\mathrel{\spose{\lower 3pt\hbox{$\mathchar"218$}}
     \raise 2.0pt\hbox{$\propto$}}}

\begin{document}

\input epsf.tex    

\input psfig.sty
\newcommand{\etal}{et al.} 
\newcommand{\radm}{rad m$^{-2}$}
\renewcommand{\deg}{\mbox{$^\circ$}}
\newcommand{\solmass}{\mbox{M$_\odot$}}
\newcommand{\peryr}{\mbox{yr$^{-1}$}} 
\newcommand\aj{{AJ}}%
\newcommand\araa{{ARA\&A}}%
\newcommand\apj{{ApJ}}%
\newcommand\apjl{{ApJ}}%
\newcommand\apjs{{ApJS}}%
\newcommand\ao{{Appl.~Opt.}}%
\newcommand\apss{{Ap\&SS}}%
\newcommand\aap{{A\&A}}%
\newcommand\aapr{{A\&A~Rev.}}%
\newcommand\aaps{{A\&AS}}%
\newcommand\azh{{AZh}}%
\newcommand\baas{{BAAS}}%
\newcommand\jrasc{{JRASC}}%
\newcommand\memras{{MmRAS}}%
\newcommand\mnras{{MNRAS}}%
\newcommand\pra{{Phys.~Rev.~A}}%
\newcommand\prb{{Phys.~Rev.~B}}%
\newcommand\prc{{Phys.~Rev.~C}}%
\newcommand\prd{{Phys.~Rev.~D}}%
\newcommand\pre{{Phys.~Rev.~E}}%
\newcommand\prl{{Phys.~Rev.~Lett.}}%
\newcommand\pasp{{PASP}}%
\newcommand\pasj{{PASJ}}%
\newcommand\qjras{{QJRAS}}%
\newcommand\skytel{{S\&T}}%
\newcommand\solphys{{Sol.~Phys.}}%
\newcommand\sovast{{Soviet~Ast.}}%
\newcommand\ssr{{Space~Sci.~Rev.}}%
\newcommand\zap{{ZAp}}%
\newcommand\nat{{Nature}}%
\newcommand\iaucirc{{IAU~Circ.}}%
\newcommand\aplett{{Astrophys.~Lett.}}%
\newcommand\apspr{{Astrophys.~Space~Phys.~Res.}}%
\newcommand\bain{{Bull.~Astron.~Inst.~Netherlands}}%
\newcommand\fcp{{Fund.~Cosmic~Phys.}}%
\newcommand\gca{{Geochim.~Cosmochim.~Acta}}%
\newcommand\grl{{Geophys.~Res.~Lett.}}%
\newcommand\jcp{{J.~Chem.~Phys.}}%
\newcommand\jgr{{J.~Geophys.~Res.}}%
\newcommand\jqsrt{{J.~Quant.~Spec.~Radiat.~Transf.}}%
\newcommand\memsai{{Mem.~Soc.~Astron.~Italiana}}%
\newcommand\nphysa{{Nucl.~Phys.~A}}%
\newcommand\physrep{{Phys.~Rep.}}%
\newcommand\physscr{{Phys.~Scr}}%
\newcommand\planss{{Planet.~Space~Sci.}}%
\newcommand\procspie{{Proc.~SPIE}}%
\let\astap=\aap
\let\apjlett=\apjl
\let\apjsupp=\apjs
\let\applopt=\ao
\newcommand\phn{\phantom{0}}%
\newcommand\phd{\phantom{.}}%
\newcommand\phs{\phantom{$-$}}%
\newcommand\phm[1]{\phantom{#1}}%
\let\la=\lesssim            
\let\ga=\gtrsim
\newcommand\sq{\mbox{\rlap{$\sqcap$}$\sqcup$}}%
\newcommand\arcdeg{\mbox{$^\circ$}}%
\newcommand\arcmin{\mbox{$^\prime$}}%
\newcommand\arcsec{\mbox{$^{\prime\prime}$}}%
\newcommand\fd{\mbox{$.\!\!^{\mathrm d}$}}%
\newcommand\fh{\mbox{$.\!\!^{\mathrm h}$}}%
\newcommand\fm{\mbox{$.\!\!^{\mathrm m}$}}%
\newcommand\fs{\mbox{$.\!\!^{\mathrm s}$}}%
\newcommand\fdg{\mbox{$.\!\!^\circ$}}%
\def\farcm@apj{%
 \mbox{.\kern -0.7ex\raisebox{.9ex}{\scriptsize$\prime$}}%
}%
\def\farcs@apj{%
 \mbox{%
  \kern  0.13ex.%
  \kern -0.95ex\raisebox{.9ex}{\scriptsize$\prime\prime$}%
  \kern -0.1ex%
 }%
}%

\jname{Annual Reviews of Astronomy and Astrophysics}
\jyear{2002}
\jvol{40}

\title{Cluster Magnetic Fields}

\markboth{Cluster Magnetic Fields}{Carilli \& Taylor}

\author{C.\L.\ Carilli \& G.\ B.\ Taylor
\affiliation{National Radio Astronomy Observatory, Socorro, NM 87801,
  USA; ccarilli@nrao.edu, gtaylor@nrao.edu}}


\begin{abstract}

  Magnetic fields in galaxy clusters have been measured using a
  variety of techniques, including: studies of synchrotron relic
  and halo radio sources within clusters, studies of inverse Compton
  X-ray emission from clusters, surveys of Faraday rotation measures
  of polarized radio sources both within and behind clusters, and
  studies of Cluster Cold Fronts in X-ray
  images.  These measurements imply that most cluster atmospheres are
  substantially magnetized, with typical field strengths of order 1
  $\mu$Gauss with high areal filling factors out to Mpc
  radii.  There is likely, however, to be considerable variation in
  field strengths and topologies both within and between clusters,
  especially when comparing dynamically relaxed clusters to those that
  have recently undergone a merger.  In some locations, such as the
  cores of cooling flow clusters, the magnetic fields reach
  levels of 10--40 $\mu$G and may be dynamically important.  In all
  clusters the magnetic fields have a significant effect on energy
  transport in the intracluster medium. We also review current
  theories on the origin of cluster magnetic fields.

\end{abstract}

\maketitle

\section{Introduction}

Magnetic fields play an important role in virtually all astrophysical
phenomena.  Close to home, the Earth has a bipolar magnetic field with
a strength of 0.3 G at the equator and 0.6 G at the poles.  This
field is thought to originate in a dynamo due to fluid motions within
the liquid core \citep{Soward83}.  With its faster angular rotation,
Jupiter leads the planets with an equatorial field strength of $\sim$
4 G \citep{Warwick63, Smith74}.  A similar mechanism produces
the solar magnetic fields which give rise to spectacular sunspots,
arches, and flares \citep{Parker79}.  
Within the interstellar medium, 
magnetic fields are thought to regulate star formation via
the ambipolar diffusion mechanism \citep{spitzer78}.
Our own Galaxy has a typical interstellar magnetic field strength
of $\sim$2 $\mu$G in a regular ordered component on kiloparsec scales,
and a similar value in a smaller-scale, random component
\citep{Beck96,Kulsrud99}.  Other spiral galaxies have been
estimated to have magnetic field strengths of 5 to 10 $\mu$G, with
fields strengths up to 50$\mu$G found in starburst galaxy nuclei
\citep{Beck96}.  Magnetic fields are
fundamental to the observed properties of jets and lobes in radio
galaxies \citep{Bridle84}, and may be primary elements in the
generation of relativistic outflows from accreting, massive black
holes \citep{begelman84}.  Assuming equipartition conditions apply,
magnetic field strengths range from a few $\mu$G in kpc-scale extended
radio lobes, to mG in pc-scale jets.

The newest area of study of cosmic magnetic fields is on larger scales
still, that of clusters of galaxies. Galaxy clusters are the largest
virialized structures in the universe. The first spatially resolving
X-ray observations of clusters \citep{Forman72} revealed atmospheres of
hot gas ($10^7$ to $10^8$ K) which extend to Mpc radii and which
dominate the baryonic mass of the systems ($10^{13}$ to 
$10^{14}$ M$_\odot$). Soon thereafter came the first attempts to
measure magnetic field strengths in the intracluster medium (ICM)
\citep{Jaffe77}. Only in the last decade has it
become clear that magnetic fields are ubiquitous in cluster
atmospheres, certainly playing a critical role in determining the
energy balance in cluster gas through their effect on heat conduction,
and in some cases perhaps even becoming important dynamically.

Cluster magnetic fields have been treated as secondary topics in 
reviews of cluster atmospheres \citep{Sarazin88,Fabian94},
and in general reviews of cosmic magnetic fields 
\citep{Kronberg96,ruzmaikin87}. To date there has been no
dedicated review on cluster magnetic fields. 

The focus of this review
is primarily observational. We summarize and critique
various methods used for measuring cluster magnetic fields. 
In the course of the review we consider important effects of
magnetic fields in clusters, such as their effect on heat conduction
and gas dynamics, and other issues such as the lifetimes of
relativistic particles in the ICM. We then attempt to synthesize
the various measurements and develop a general picture for
cluster magnetic fields, with the caveat that there may be
significant differences between clusters,
and even within a given cluster atmosphere. We conclude with a
section on the possible origin of cluster magnetic fields.

We assume H$_0 = 75$ km s$^{-1}$ Mpc$^{-1}$ and q$_0$=0.5, unless
stated otherwise. Spectral index, $\alpha$, is defined
as $S_\nu \propto \nu^\alpha$. 

\section{Synchrotron radiation}

\subsection{Radio halos} 

\begin{figure}[htp]
\centerline{\epsfxsize=5.3in\epsffile{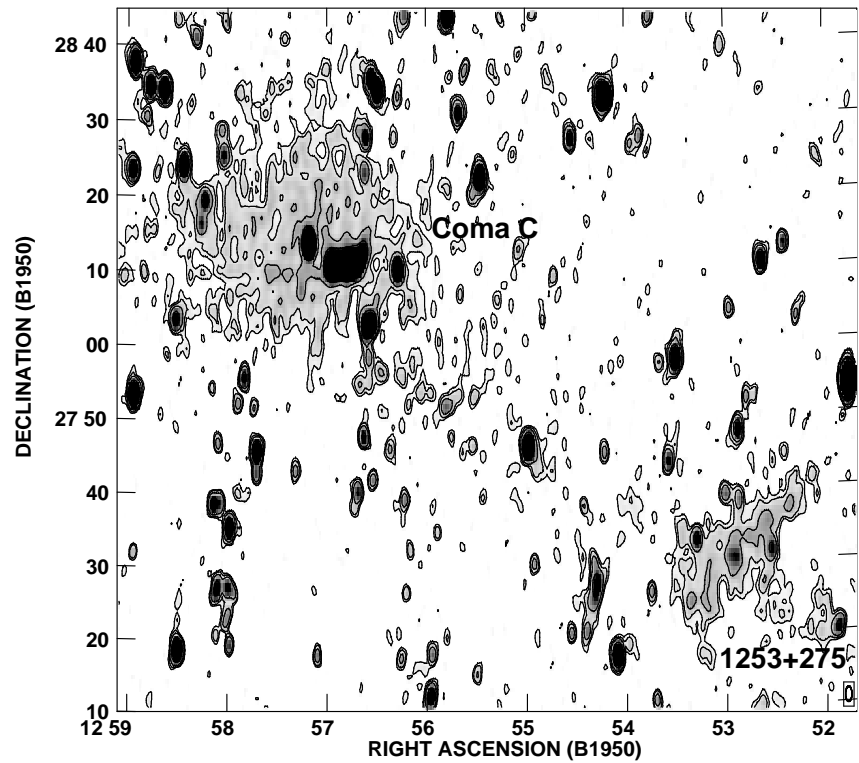}}
\caption{\footnotesize WSRT radio image of the Coma cluster region at 90 cm,
with angular resolution of 55$^{\prime\prime}$ $\times$
125$^{\prime\prime}$ (HPBW, RA $\times$ DEC) from Feretti et al (1998).
Labels refer to the halo source Coma C and the relic source 1253+275.
The grey scale range displays total intensity emission from 2 to 
30 mJy/beam while contour levels
are drawn at 3, 5, 10, 30, and 50 mJy/beam. The bridge of
radio emission connecting Coma C to 1253+275 is resolved and visible 
only as a region with an apparent higher
positive noise.
The Coma cluster is at a redshift of 0.023, such that $1' = 27$ kpc.
\label{coma}}
\end{figure}

   Over 40 years ago \cite{Large59} discovered a radio source in the Coma
cluster that was extended even when observed with a 45\arcmin\ beam.
This source (Coma C) was studied by \cite{Willson70} who found
that it had a steep spectral index and could not be made up of
discrete sources, but instead was a smooth ``radio halo'' with no
structure on scales less than 30\arcmin.  Willson further surmised
that the emission mechanism was likely to be synchrotron, and if in
equipartition required a magnetic field strength of 2 $\mu$G.  
In Fig.~\ref{coma} we show
the best image yet obtained of the radio halo in the Coma
cluster. Other
radio halos were subsequently discovered, although the number known
remained under a dozen until the mid-90s 
\citep{Hanisch82}.  

Using the Northern VLA Sky Survey (NVSS; \cite{Condon98}) and X-ray
selected samples as starting points
\cite{Giovannini00, Giovannini99} have performed moderately deep VLA
observations (integrations of a few hours) which have 
more than  doubled the
number of known radio halo sources.   Several
new radio halos have also been identified from the Westerbork
Northern Sky Survey \citep{Kempner01}.
These radio halos typically have sizes 
$\sim$1 Mpc, steep spectral indices ($\alpha < -1$), low fractional
polarizations ($< 5\%$), low surface
brightnesses ($\sim 10^{-6}$ Jy arcsec$^{-2}$ at 1.4 GHz), 
and centroids close to the cluster center defined 
by the  X-ray emission.

A steep correlation between cluster X-ray and
radio halo luminosity has been found,  as well as a correlation
between radio and X-ray surface brightnesses in clusters 
\citep{liang00,feretti01,govoni01}.
A complete (flux limited) sample of X-ray clusters shows only
5$\%$ to 9$\%$ of the sources are
detected at the surface brightness limits of 
the NVSS of 2.3 mJy beam$^{-1}$, where the beam has FWHM = 45\arcsec 
\citep{Giovannini00,feretti01}.  But this sample contains mostly 
clusters with X-ray luminosities $< 10^{45}$ erg s$^{-1}$.  If one
selects for clusters with X-ray luminosities $> 10^{45}$ erg
s$^{-1}$, the radio detection rate increases to 35$\%$
\citep{feretti01,owen99}.  Likewise, there may be a correlation between
the existence of a cluster radio halo and the existence of
substructure in X-ray images of the hot cluster atmosphere, indicative
of  merging clusters, and a corresponding anti-correlation between
cluster radio halos and clusters with relaxed morphologies,
{\it e.g.}, cooling flows   \citep{govoni01}, although
these correlations are just beginning to be quantified 
\citep{Buote01}.

Magnetic fields in cluster radio halos can be derived 
assuming a minimum energy configuration for the
summed energy in relativistic particles and magnetic fields
\citep{burbidge59}, corresponding roughly to energy equipartition
between fields and 
particles. The equations for deriving minimum energy fields from 
radio observations are given in \cite{miley80}.
Estimates for minimum energy magnetic field strengths in
cluster halos range from  0.1 to 1 $\mu$G \citep{Feretti99b}.
One of the  best studied halos is that in Coma, 
for which \cite{Giovannini93}
report a minimum energy  magnetic field of 0.4 $\mu$G.
These calculations typically assume $k = 1$, $\eta = 1$, 
$\rm \nu_{low} = 10$ MHz, and $\rm \nu_{high} = 10$ GHz, 
where $k$  is the ratio of energy densities in 
relativistic protons to that in electrons, $\eta$ is the volume
filling factor, $\rm \nu_{low}$ is the low frequency cut-off for the
integral, and $\rm \nu_{high}$ is the high frequency
cut-off.  All of these parameters are poorly constrained,
although the magnetic field strength only behaves as these parameters
raised to the  ${{2}\over{7}}$ power. For example, using a value of
$k \sim 50$, as observed for Galactic cosmic rays \citep{meyer69},
increases the fields by a factor of three. 

\cite{Brunetti01a} present a method for estimating
magnetic fields in the Coma cluster radio halo independent of
minimum energy assumptions. They base their analysis on 
considerations of the observed radio and X-ray spectra, the
electron inverse Compton and synchrotron radiative
lifetimes, and reasonable mechanisms for 
particle reacceleration. They conclude that the fields
vary smoothly from 2$\pm$1 $\mu$G in the cluster center,
to 0.3$\pm$0.1 $\mu$G at 1 Mpc radius. 

\subsection{Radio relics} 

A possibly related phenomena to radio halos is a class of sources
found in the outskirts of clusters known as radio relics.  Like the
radio halos, these are very extended sources without an identifiable
host galaxy (Fig.~\ref{coma}).  Unlike radio halos, radio
relics are often 
elongated or irregular in shape, are located at the cluster periphery
(by definition), and are strongly polarized, up to 50\% in the case of
the relic 0917+75 \citep{Harris93}.  As the name implies, one of the
first explanations put forth to explain these objects was that these
are the remnants of a radio jet associated with an 
active galactic nucleus (AGN) that has since turned off and
moved on.  A problem with this model is that, once the energy source
is removed,  the radio source is
expected to fade on a timescale $<< 10^8$ years due to
adiabatic expansion, inverse Compton, and synchrotron losses
(see \S 4.1). This short timescale
precludes significant motion of the host galaxy 
from the vicinity of the radio source.

A more compelling explanation is that the relics are
the result of first order Fermi acceleration (Fermi I) of relativistic 
particles in shocks produced during cluster mergers \citep{Ensslin98},
or are fossil radio sources revived by compression associated with
cluster mergers \citep{Ensslin01}.
Equipartition field strengths for relics range from 0.4 -- 2.7
$h_{50}^{2/7} \mu$G \citep{Ensslin98}.  If the relics are
produced by shocks or compression during a cluster
merger, then \cite{Ensslin98} calculate a pre-shock cluster
magnetic field strength in the range 0.2-0.5 $\mu$G.

\section{Faraday rotation}
\subsection{Cluster center sources}

The presence of a magnetic field in an ionized plasma sets a
preferential direction for the gyration of electrons, leading to
a difference in the index of refraction for 
left versus right circularly polarized  radiation.  
Linearly polarized light 
propagating through a magnetized plasma
experiences a phase shift of the left versus right circularly
polarized components of the wavefront, leading to a 
rotation of the plane of polarization, 
$\Delta\chi$ = RM $\lambda^2$, where $\Delta\chi$ is the change
in the position angle of polarization, 
$\lambda$ is the wavelength of the radiation,
and RM is the Faraday rotation measure. 
The RM is related to the electron density, $n_{\rm
  e}$, and the magnetic field, $\bf B$, as:
$${\rm RM} = 812\int\limits_0^L n_{\rm e} {\bf B} \cdot d{\bf l} 
~{\rm  radians~m}^{-2}~,
\eqno(1)
$$                    
where ${\bf B}$ is measured in $\mu$Gauss,  $n_{\rm e}$
in cm$^{-3}$ ~and $d${\bf l} in kpc, and the bold face symbols
represent the vector product between the magnetic field and the 
direction of propagation. 
This phenomenon can also be understood qualitatively by considering
the forces on the electrons. 

Synchrotron radiation from cosmic radio sources is well known to be
linearly polarized, with fractional polarizations up to 70$\%$ in some
cases \citep{Pacholczyk70}.  Rotation measures can be derived from
multifrequency polarimetric observations of these sources by measuring
the position angle of the polarized radiation as a function of
frequency.  The RM values can then be combined with measurements of
$n_{\rm e}$ to estimate the magnetic fields.  Due to the
vector product in Eq.~1, only the magnetic field component along
the line-of-sight is measured, so the results depend on the assumed
magnetic field topology.

Most extragalactic radio sources exhibit Faraday rotation
measures (RMs) of the order of 10's of \radm\ due to propagation
of the emission through the interstellar
medium of our galaxy \citep{Simard81}.  Sources at Galactic latitudes 
$\le 5^o$ can exhibit $\sim$300 \radm.  For the past 30 years,
however, a small number 
of extragalactic sources were known to have far higher RMs than could
be readily explained as Galactic in origin.  Large intrinsic RMs were
suspected, but the mechanism(s) producing them were unclear.

\begin{figure}[t]
\centerline{\epsfxsize=5.3in\epsffile{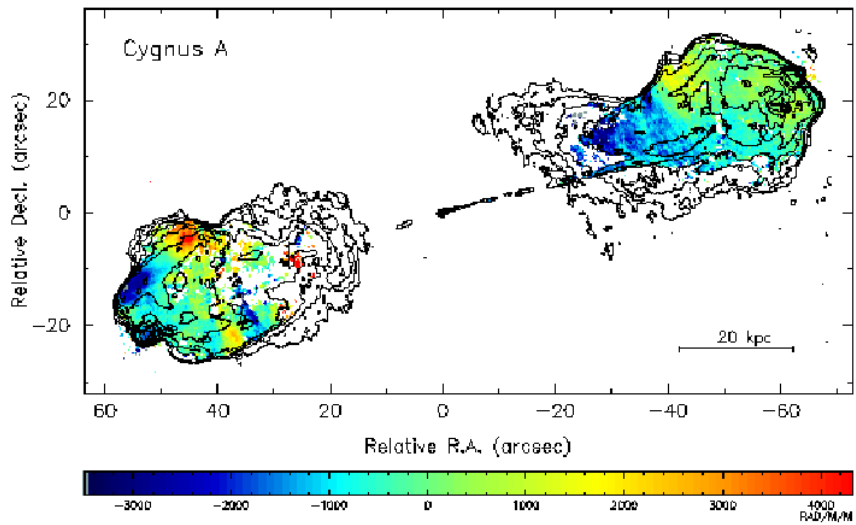}}
\caption{\footnotesize The RM distribution in Cygnus~A based on 
multi-frequency,
multi-configuration VLA
observations. The resolution is 0.35\arcsec\ (Dreher et al. 1987).  The
colorbar indicates the range in RMs from $-$3400 to $+$4300 \radm.  Note
the undulations in RM on scales of 10--30 kpc.  Contours are overlaid
from a 5 GHz total intensity image.
The RM was solved for by fitting for the change in
polarization angle with frequency on a pixel-by-pixel basis (see 
Fig.~5).
}  
\label{cyga}
\end{figure}

\cite{mitton71} discovered that the powerful radio galaxy Cygnus~A had
large, and very different RMs (35 vs $-$1350 \radm), in its two lobes.
While its low galactic latitude (5.8\deg) could possibly be invoked
to explain the high RMs, the large difference in RMs over just 
2\arcmin\ was difficult to reproduce in the context of Galactic models
\citep{alexander84}.
This ``RM anomaly'' was clarified when \cite{Dreher87} performed the
first high resolution RM studies with the VLA and found complex
structure in the RM distribution on arcsec scales (Fig.~\ref{cyga}),
with gradients as large as 600 rad m$^{-2}$ arcsec$^{-1}$. 
These large gradients conclusively ruled out a Galactic origin
for the large RMs.  

Perhaps just as important as the observed RM structure across
the lobes of Cygnus A was the discovery that the 
observed position angles behave quadratically with wavelength to
within very small errors over a wide range in wavelengths
\citep{Dreher87}. Examples of this
phenomenon are shown in (Fig.~\ref{hydfits}). Moreover,
the change in position angle from short to long wavelengths is
much larger than $\pi$ radians in many cases, while the 
fractional polarization remains constant.  This result is critical
for interpreting the large RMs for cluster center
radio sources, providing rigorous proof that the large RMs cannot be
due to  thermal gas mixed with the radio emitting plasma
\citep{Dreher87}. Such mixing would 
lead to rapid depolarization with increasing wavelength, and
departures from a quadratic behavior of $\chi$ with wavelength
\citep{burn66}. 

The Cygnus A  observations were the first to show  that the
large  RMs must arise in an external screen of magnetized, ionized
plasma, but cannot be Galactic in origin. 
\cite{Dreher87} considered a number of options for the Faraday
screen toward Cygnus A, and concluded that the most likely
site was the X-ray emitting cluster atmosphere enveloping the
radio source \citep{Fabbiano79}. They found that 
magnetic fields in the cluster gas of
2--10 $\mu$G could produce the observed RMs.

Since the ground-breaking observations of Cygnus A, RM studies of
cluster center radio sources have become a standard tool for
measuring cluster fields. RM studies of radio galaxies in clusters can
be  divided into studies of cooling-flow and non-cooling-flow
clusters. Cooling-flow clusters are those in which the X-ray emission
is strongly peaked at the center, leading to high densities and
cooling times of the hot ICM in the inner $\sim$100 kpc of 
much less than the Hubble time.  To maintain hydrostatic equilibrium,
an inward flow may be required \citep{Fabian91}.
Typical mass cooling flow rates are 100 \solmass~\peryr.  The actual
presence of material ``cooling'' and ``flowing'' is a topic
that is hotly debated at present \citep{binney01}.
What is more agreed upon is that cooling-flow clusters are more
dynamically relaxed than non-cooling flow clusters
which often show evidence of cluster mergers \citep{Markovic01}.

\begin{figure}[t!]
\centerline{\epsfxsize=5.0in\epsffile{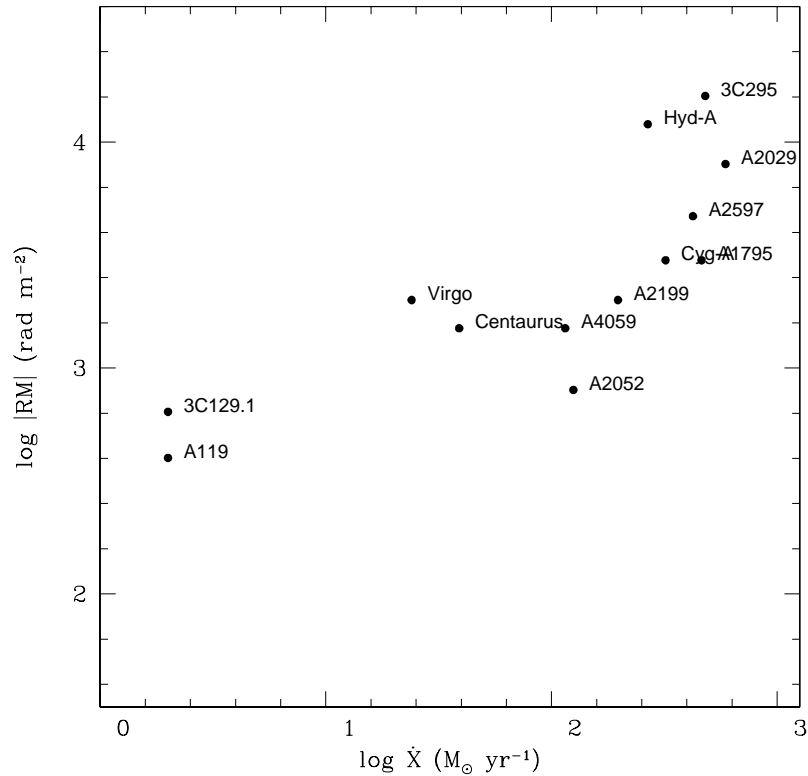}}
\caption{\footnotesize The maximum absolute RM plotted as a function of
the estimated cooling flow rate, $\dot{X}$, for a sample of X-ray 
luminous clusters with measured RMs
from Taylor et al. (2001b).  Both RM and $\dot{X}$ are expected to depend
on density to a positive power, so in that sense the correlation is
expected.}  
\label{rmxdot}
\end{figure}

Radio galaxies in cooling flow clusters attracted some of the first detailed
RM studies by virtue of their anomalously high RMs ({\it e.g.}, A1795:
\citep{Ge93}; Hydra A: \citep{Taylor93}).  Out of a sample of 14
cooling-flow clusters with strong embedded radio sources
\cite{Taylor94,Taylor01b}, found that 10 of 14 sources
display RMs in excess of 800 \radm, two (PKS\,0745{\tt -}191 and 3C\,84
in Abell 426) could not be measured due to a lack of polarized flux, and
two (Abell 119 \citep{Feretti99} and 3C\,129\citep{Taylor01a}) have lower RMs,
but with better X-ray observations turn out not to be in cooling-flow
clusters.  Hence, current data are consistent with all radio galaxies
at the center of cooling flow clusters having extreme RMs, with the
magnitude of the RMs roughly proportional to the cooling flow rate
(see Fig.~\ref{rmxdot}).
 
\begin{figure}[t!]
\centerline{\epsfxsize=5.3in\epsffile{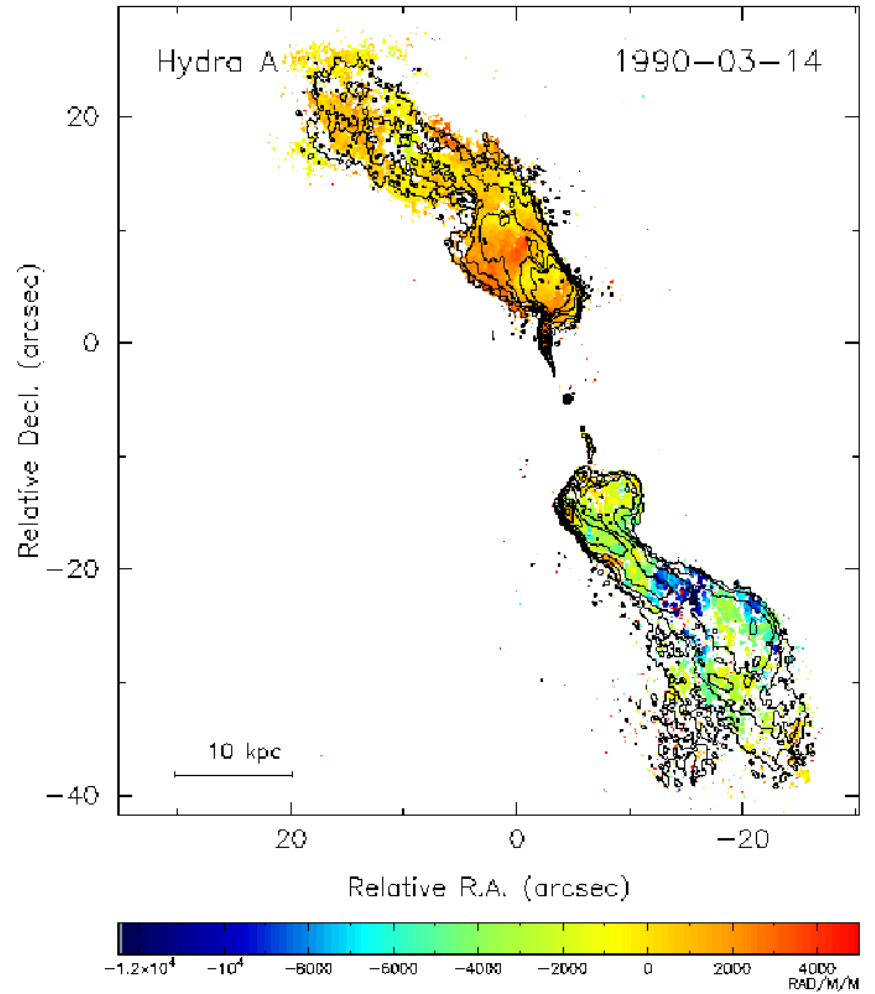}}
\caption{\footnotesize The RM distribution in Hydra~A at a 
resolution of 0.3\arcsec\ 
(Taylor \& Perley 1993) with total intensity contours overlaid. 
Multi-configuration VLA observations
were taken at 4 widely spaced 
frequencies within the 8.4 GHz band, and a single frequency in the
15 GHz band.  The
colorbar indicates the range in RMs from $-$12000 to $+$5000 \radm.
}  
\label{hyda}
\end{figure}

\begin{figure}[t!]
\centerline{\psfig{figure=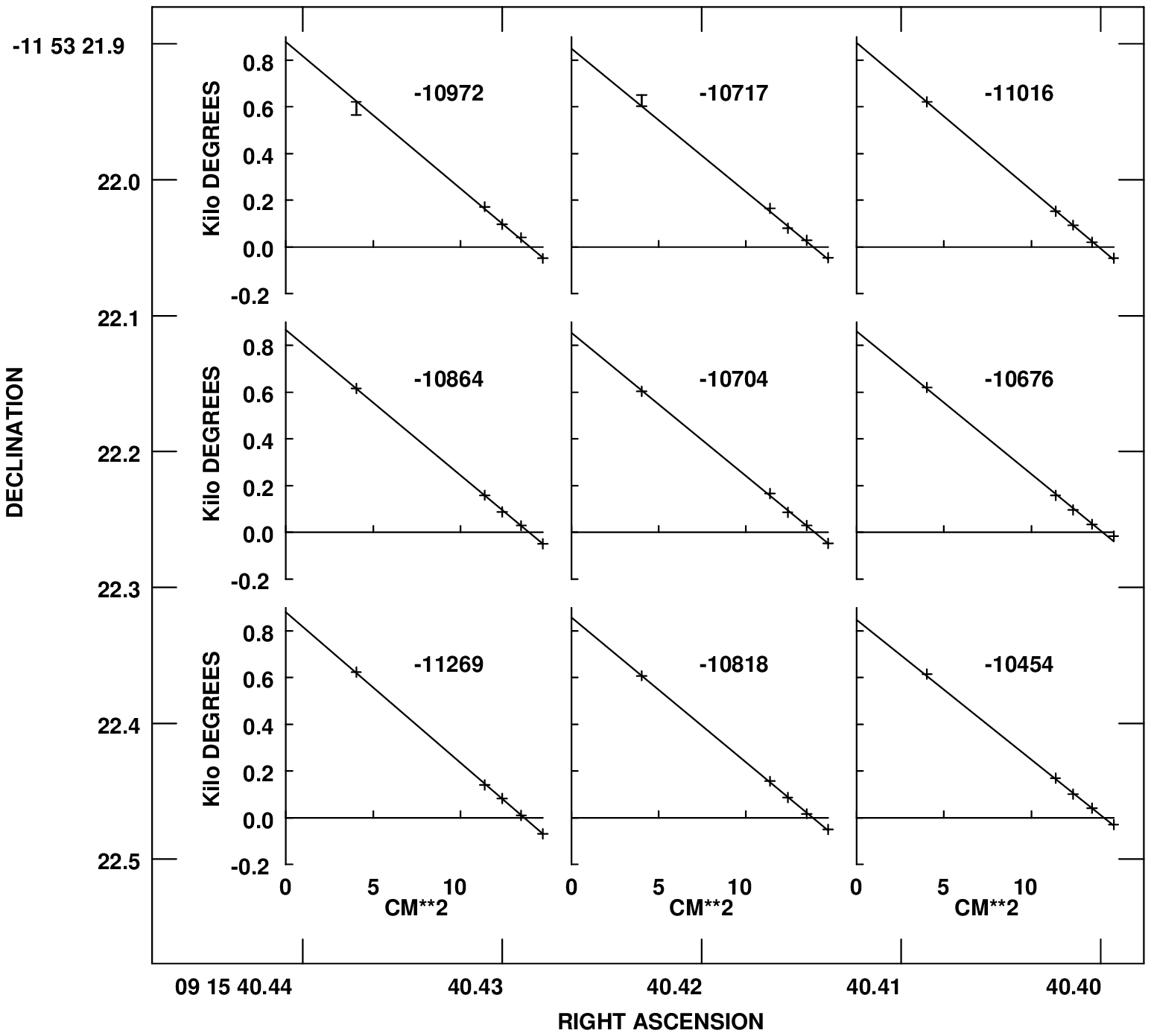,width=5.2in}}
\caption{\footnotesize The observed position angles, $\chi$, of the
linearly polarized radio emission as 
a function of the square of the observing wavelength, $\lambda^2$,
for a number of positions in the southern lobe of Hydra~A at a
resolution of 0.3\arcsec\ (Taylor \& Perley 1993).  The points plotted
are each separated by approximately one beamwidth and thus
are independent of each other.  This illustrates the consistency
of the RMs within a coherence length of $\sim$7 kpc.  Notice
also the excellent agreement to a $\lambda^2$-law for
$\Delta\chi$ = 600 degrees, nearly two complete turns.}  
\label{hydfits}
\end{figure}

The RM distributions for radio sources found at the centers of cooling
flow clusters tend to be patchy with coherence lengths of 5--10 kpc
(Fig.~\ref{hyda}). Larger ``patches'' up to 30 kpc are seen for example in
Cygnus~A (Fig.~\ref{cyga}).   In both Cygnus~A and Hydra~A one can find 
``bands'' of alternating high and low RM (see Figures \ref{cyga} and \ref{hyda}).  Such bands
are also found in the non-cooling flow cluster sources \citep{Eilek01},
along with slightly larger coherence lengths of 15--30 kpc.
In Hydra~A there is a strong trend for
all the RMs to the north of the nucleus to be positive and to the south
negative.  To explain this requires a field reversal
and implies a large-scale (100 kpc) ordered
component to the cluster magnetic fields in Hydra~A.  \cite{Taylor93}
found the large scale field strength to be $\sim$7 $\mu$G, and 
more tangled fields to have a strength of $\sim$40 $\mu$G.  A similar RM
sign reversal across the nucleus is seen in A1795, although in this
case the radio source is only 11 kpc in extent. 

Minimum cluster magnetic field strengths can be estimated by 
assuming a constant magnetic field along the line-of-sight
through the cluster.  Such estimates usually lead to magnetic
field strengths of 5 to 10 $\mu$G in cooling flow clusters,
and a bit less (factor $\sim$2) in the non-cooling flow clusters.

From the patchiness of the RM distributions it is clear that
the magnetic fields are not regularly ordered on cluster (Mpc)
scales, but have coherence scales between 5 and 10 kpc.
Beyond measuring a mean line-of-site field, the next level
of RM modeling entails `cells' of constant size
and magnetic field strength, but random magnetic field
direction, uniformly filling the cluster.  The  RM
produced by such a screen will be built up in a random walk fashion
and will thus have an average value of 0 \radm, but a dispersion in
the RM, $\sigma_{\rm RM}$, that is proportional to the square root of
the number of cells along the line-of-sight through the cluster.  
The most commonly-fit form to the X-ray observations to obtain 
the radial electron density distribution,
$n_e(r)$ through a cluster is the modified King model \citep{Cavaliere76}:

$$
n_e(r) =n_0 (1 + r^2/r^2_{\rm c})^{-3 \beta/2},
\eqno(2)
$$

\noindent where $n_0$ is the central density, $r_c$ is the core
radius, and $\beta$ is a free parameter in the fit.
Typical values for these parameters are $r_c \sim 200$ kpc,
$\beta \sim {{2}\over{3}}$, and $n_o \sim 0.01$ cm$^{-3}$. 

For this density profile and cells of constant
magnetic strength but random orientation, \cite{Felten96, Feretti95}
derived the following relation for the RM dispersion:

$$
\sigma_{\rm RM} = {{{\rm K} B~n_0~r_c^{1/2} l^{1/2}}
\over{(1 + r^2/r_c^2)^{(6\beta - 1)/4}}}
\sqrt{{\Gamma(3\beta - 0.5)}\over{{\Gamma(3\beta)}}}
\eqno(3)
$$

\noindent where $l$ is the cell size, $r$ is the distance of           
the radio source from the cluster center, $\Gamma$ is the Gamma
function,
and K is a factor which depends on the location of the radio
source along the line-of-sight through the cluster: K = 624
if the source is beyond the cluster, and K = 441 if the source
is halfway through the cluster. Note that Eq.~3
assumes that the magnetic
field strength, $B$, is related to the component along the
line of sight, ($B_{\|}$), by $B = \sqrt{3} B_\|$. The cell size, $l$, can be
estimated to first order from the observed coherence lengths
of the RM distributions.  Both cooling-flow and non-cooling
flow clusters yield typical estimates of 5 to 10 kpc.
Magnetic field strength estimates, however are two to three
times higher in the cooling-flow clusters -- 19 $\mu$G
in the 3C\,295 cluster \citep{Allen01a} compared to 6 $\mu$G in
the 3C\,129 cluster \citep{Taylor01a} using the methodology
described above.  

Most radio sources found embedded in clusters are located at the
center and identified with a cD galaxy.  This relatively high 
pressure environment has been found in many cases to confine or
distort the radio galaxy \citep{Taylor94}, as well as giving rise to
extreme RMs.  For this same reason the extended radio sources in 
Hydra~A and Cygnus~A are unique in that they sample regions 
over 100 kpc in linear extent.  There are, however, a few
clusters containing more  
than one strong, polarized radio source.  The cluster 
Abell 119 \citep{Feretti99} contains three radio galaxies.  
Using an analysis based on Eq.~3 above, \cite{Feretti99} find  that a
magnetic field strength of 6--12 $\mu$G extending
over 3 Mpc could explain the RM distributions for all 3 sources, although
they note that such a field would exceed the thermal pressure in
the outer parts of the cluster.

In a reanalysis of the Abell 119 measurements,  
\cite{Dolag01} find that the 
field scales as $n_e^{0.9}$. This power-law behavior is
marginally steeper than that expected assuming flux conservation, for
which the tangled field scales as $n_e^{2/3}$, and significantly
steeper than that expected assuming a constant ratio between magnetic
and thermal energy  density, for which 
the tangled field scales as $n_e^{1/2}$ for an isothermal atmosphere.
In the 3C\,129 cluster there are two extended radio galaxies
whose RM observations can be fit by a field strength of 6 $\mu$G.
Finally in A514, \cite{Govoni01} has measured the RM distributions 
of 5 embedded (and background) radio sources and found
cluster magnetic field strengths of 4--9 $\mu$G spread over
the central 1.4 Mpc of the cluster. If the magnetic field
scales with the density raised to a positive power, then the 
product  of $B$ and $n_e$ in Eq.~1 implies that the observed
rotation measures are heavily weighted by the innermost cells in
the cluster \citep{Dreher87}. 

It has been suggested that high RMs may  result from an
interaction between the radio galaxy and the ICM, such that the RMs
are generated locally, and are not indicative of cluster magnetic
fields.  \cite{Bicknell90} proposed a model in which the RM 
screen is due to a 
boundary layer surrounding the radio source in 
which the large magnetic fields within the radio source 
are mixed with the large thermal densities outside the radio source
by Kelvin-Helmholtz waves along the contact discontinuity. 
This model predicts a
Faraday depolarized region of a few kpc extent
surrounding the radio source, where the
synchrotron emitting material has mixed with the thermal gas. 
Such a depolarized shell has not been observed to date. 

In general, extreme RMs have been observed in sources of very
different morphologies, from edge-brightened (Fanaroff-Riley Class II
\citep{Fanaroff74}), to edge darkened (FR I) sources. The models for
the hydrodynamic evolution of these different classes of sources are
thought to be very different, with the FRII sources expanding
supersonically, while the FRI sources expand sub-sonically
\citep{begelman84}.  This argues that the high RMs are not solely a
phenomenon arising due to a local interaction between the radio source
and its environment, but are more likely to be a property of the large
scale ({\it i.e.}, cluster) environment.  Perhaps the most telling
argument against the interaction model is that RM studies of
background radio sources seen through cluster atmospheres also
indicate $\mu$G cluster magnetic fields (see \S 3.2).

While we feel that large RMs for cluster center radio sources most
likely arise in the large scale cluster atmosphere, 
we should point out that there are some cases in which the radio
source does appear to compress 
the gas and fields in the ICM to produce local RM enhancements. For
example, there is evidence for an RM enhancement due to the bow shock
preceding the radio hot spots in Cygnus A and 3C\,194
\citep{Carilli88,Taylor92}.  However, even in these cases the implied
external ({\it i.e.}, unperturbed) ICM fields are a few $\mu$G.

\subsection{Background sources}

The first successful demonstration of Faraday rotation 
of the polarized emission from  background radio 
sources seen through a cluster atmosphere was presented by
\cite{Vallee86} for A2319.  \cite{Vallee87} combined the RM
excess in A2319 with density estimates from X-ray observations
by \cite{Jones79} to estimate a field strength of 2 $\mu$G if the field
is organized in cells of size 20 kpc.  
\cite{Kim90} found that the RMs toward sources within 20\arcmin\ of the Coma
cluster center had an enhanced RM dispersion by 38 $\pm$ 6 \radm.
From this excess they derived a magnetic field strength of 2.0 $\pm$
1.1 $\mu$G assuming a cell size in the range 7--26 kpc.
\cite{Feretti95} found evidence from the RM distribution of the
embedded cluster source NGC\,4869 for
smaller cell sizes ($\sim$1 kpc), and subsequently estimated the field
strength in Coma to be 6.2 $\mu$G.

\begin{figure}[t!]
\centerline{\epsfxsize=4.2in\epsffile{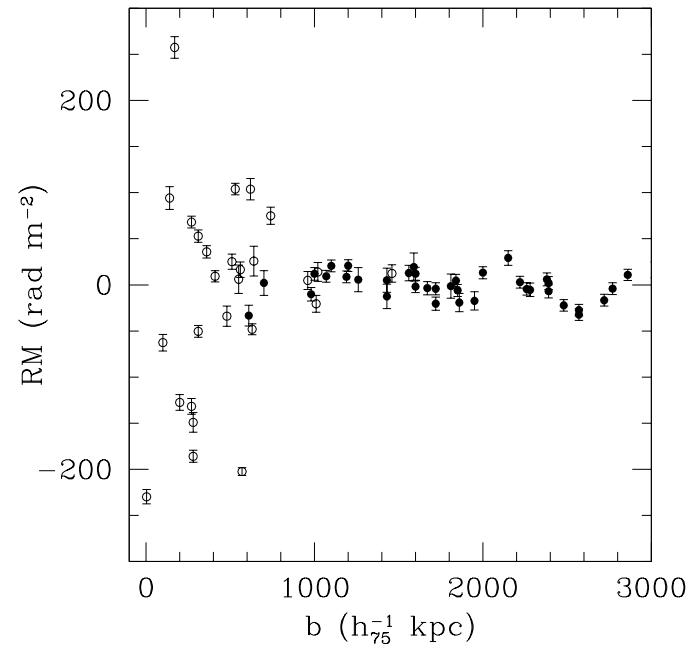}}
\caption{\footnotesize The integrated RM plotted as a function of source impact
parameter in kiloparsecs for the sample of 16 Abell clusters described
in Clarke et al (2001).  The open symbols represent sources
viewed through the cluster, while the closed symbols represent the control
sample of field sources.}  
\label{rrm}
\end{figure}

The most significant work in this area is the recent VLA survey by
Clarke et al. (2001), in which they observed radio sources in and
behind a representative sample of 16 Abell clusters at $z < 0.1$.
They find enhanced rotation measures on the large majority of the
lines of sight within 0.5 Mpc of the cluster centers (Fig.~\ref{rrm}),
from which they derive an areal filling of the magnetic fields of
95$\%$.  Their modeling results in magnetic fields of $\sim$5 $\mu$G,
assuming a cell size of 10 kpc.  These clusters were chosen not to
have cooling flows, but are otherwise unremarkable in their
properties.  By observing sources behind the clusters, these
observations demonstrate that an embedded powerful radio galaxy is not
required to produce significant RMs.  Another advantage of this
technique is that it permits estimation of the spatial extent of the
magnetic fields within the cluster ($\sim 0.5$ Mpc). The areal filling
factor of 95$\%$ (assuming constant magnetic fields in and among all
clusters) suggests a relatively large volume filling factor for the
fields, with a formal (extreme) lower limit being about 8$\%$ for 10
kpc cell sizes.

\subsection{High redshift sources}

Radio galaxies and radio loud quasars have been detected to 
$z = 5.2$ \citep{vanbreugel00}. 
The extended polarized emission from these sources
provides an ideal probe of their environments through 
Faraday rotation observations.  Extensive radio imaging 
surveys of $z > 2$ radio galaxies and quasars have shown
large rotation measures, and Faraday depolarization, in at
least 30$\%$ of the sources,
indicating that the sources are situated behind dense Faraday
screens of magnetized, ionized plasma
\citep{chambers89,Garrington88,carilli94,carilli97,pentericci01,lonsdale,ramana},
with a possible increase in this fraction with 
redshift \citep{pentericci01}.
Drawing the analogy to lower $z$ radio galaxies, these authors 
proposed that the high $z$ sources may be embedded in magnetized 
(proto-) cluster atmospheres, with $\mu$G field strengths.

A difficulty with the study of high redshift sources is that the
sources are typically small ($<$ few arcseconds), requiring higher
frequency observations (5 to 8 GHz) in order to properly resolve the
source structure. This leads to two problems. First is that the rest
frame frequencies are then $\ge 20$ GHz, such that only extreme values
of Faraday rotation can be measured (RM $\ge 1000$ \radm).  
And second is that only the
flatter spectrum, higher surface brightness radio emitting structures
in the sources are detected, thereby allowing for only a few
lines-of-site through the ICM as RM probes.  Imaging at frequencies of
1.4 GHz or lower with sub-arcsecond resolution is required to address
this interesting issue.

\section{Inverse Compton X-ray emission}

Cosmic magnetic fields can  be derived by comparing  inverse
Compton X-ray emission and radio synchrotron radiation
\citep{harris79,rephaeli87}. Inverse 
Compton (IC) emission is the  relativisitic
extrapolation of the  Sunyaev-Zel'dovich effect \citep{rephaeli95},
involving up-scattering of the ambient photon field by
the relativisitic particle population. 
The IC process involves two Lorentz transforms (to and from the
rest frame of the electron), plus Thompson
scattering in the rest frame of the electron, leading to 
$\rm \nu_{IC} \sim {4\over 3} \gamma^2 \nu_{bg}$, where
$\nu_{\rm IC}$ is the emergent frequency of the scattered radiation,
$\gamma$ is the electron Lorentz factor, and $\nu_{\rm bg}$ is the 
incident photon frequency \citep{bagchi99}. From a quantum mechanical
perspective, synchrotron radiation is directly
analogous to IC emission, with synchrotron radiation being the
up-scattering of the virtual photons that constitute the static
magnetic field. Given a relativistic electron population, the 
IC emissivity is directly proportional to the energy density in the 
photon field, $U_{\rm bg}$,
while the synchrotron emissivity is proportional to the
energy density in the magnetic field, $U_{\rm B}$, leading to a simple
proportionality between synchrotron and IC luminosity:
$ {{L_{\rm syn}}\over{L_{\rm IC}}} \propto {{U_{\rm B}}\over{U_{\rm bg}}}$.
Given that they originate from the same (assumed power-law)
relativistic electron population, 
IC X-rays and synchrotron radio emission share the same 
spectral index, $\alpha$.
The spectral index relates to the index for the power-law
electron energy distribution, $\Gamma$, as $\Gamma = 2\alpha - 1$,
and to the photon index as $\alpha - 1$. 

In most astrophysical circumstances $U_{\rm bg}$
is dominated by the cosmic microwave background (CMB), except in
the immediate vicinity of active star forming regions and AGN
\citep{Brunetti01a,carilli01}. The
Planck function at T = 2.73 K peaks near a frequency 
of  $\nu_{\rm bg} \sim 1.6\times10^{11}$ Hz, hence
IC X-rays observed at 20 keV
($\nu_{\rm IC} = 4.8\times10^{18}$ Hz), are emitted
predominantly by electrons at $\gamma \sim 5000$, independent of
redshift\footnote{$\gamma$ is independent of redshift since $\nu_{\rm
    bg}$ increases as $1 + z$.}.   
The corresponding radio synchrotron emission 
from $\gamma = 5000$ electrons peaks at a (rest frame) frequency of
$\nu_{\rm syn} \sim 4.2 ({B\over{1 \mu G}}) \gamma^2$ Hz = 100 MHz
\citep{bagchi99}. 

Many authors have considered the problem of
deriving magnetic fields by comparing synchrotron radio and inverse
Compton X-ray emission
\citep{blumenthal70,harris79,rephaeli87,feigelson95,bagchi99}. 
Assuming $\alpha = -1$, the magnetic field is given by:

$$ B = 1.7~ (1+{\sl z})^2 ~ ({{S_{\rm r} \nu_{\rm r}}\over{S_{\rm x} \nu_{\rm x}}})^{0.5}
~~\mu {\rm G} \eqno(4) $$

\noindent where $S_{\rm r}$ and $S_{\rm x}$ are the radio and X-ray
flux densities at observed frequencies $\nu_{\rm r}$ and $\nu_{\rm x}$,
respectively.  Note that, unlike Faraday rotation measurements, 
the geometry of the field does not play a critical 
role in this calculation, except in the context of the
electron pitch angle distribution (see \S 4.2).

The principle difficulty in studying IC emission from clusters of
galaxies is confusion due to the thermal emission from the cluster
atmosphere.  One means of separating the two emission mechanisms is
through spectroscopic X-ray observations at high energy. The IC
emission will have a harder, power-law spectrum relative to thermal
brehmstrahlung emission.  Recent high energy X-ray satellites such as
Beppo/Sax and RXTE have revolutionized this field by allowing for
sensitive observations to be made at energies well above 10 keV
\citep{rephaeli01}.  Prior to these instruments, most studies of IC
emission from clusters with radio halos only provided lower limits to
the magnetic fields of about 0.1 $\mu$G \citep{rephaeli87}.

\begin{figure}[t!]
\centerline{\psfig{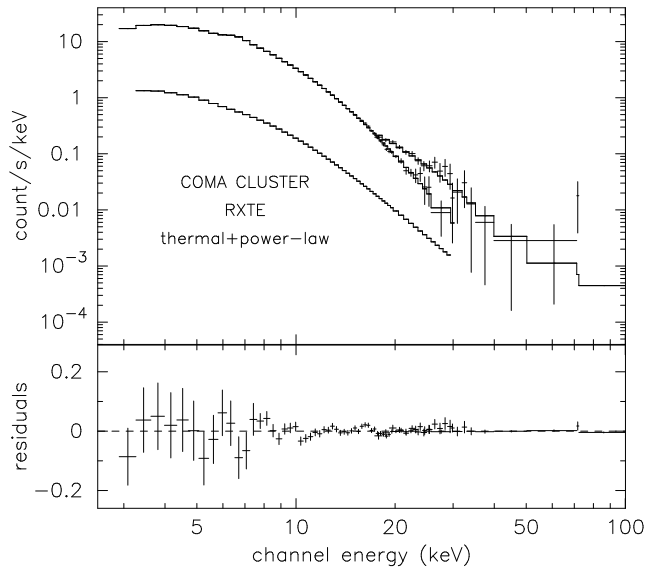}}
\caption{\footnotesize RXTE spectrum of the Coma cluster. Data and folded Raymond-Smith 
($kT \simeq 7.51$ keV), and power-law (photon index $=2.34$) models
are shown in  
the upper frame; the latter component is also shown separately in the lower 
line. Residuals of the fit are shown in the lower frame (Rephaeli 2001).
}  
\label{rxtecoma}
\end{figure}

Recent observations of four clusters with radio halos
with Beppo/Sax and RXTE have revealed hard X-ray
tails which dominate the integrated emission above 20 keV 
\citep{rephaeli99,fusco01,fusco00}.  In Fig.~\ref{rxtecoma}
we reproduce the RXTE observations of the Coma cluster.  The
hard X-ray emission in these sources has a spectral index $\alpha =
-1.3 \pm 0.3$, roughly consistent with the radio spectral
index.  Comparing the IC X-ray and radio synchrotron
emission in these sources leads to a volume averaged cluster magnetic
field of 0.2 to 0.4 $\mu$G, with a relativistic electron energy
density $\sim 10^{-13}$ erg cm$^{-3}$.

\begin{figure}[htp]
\centerline{\psfig{figure=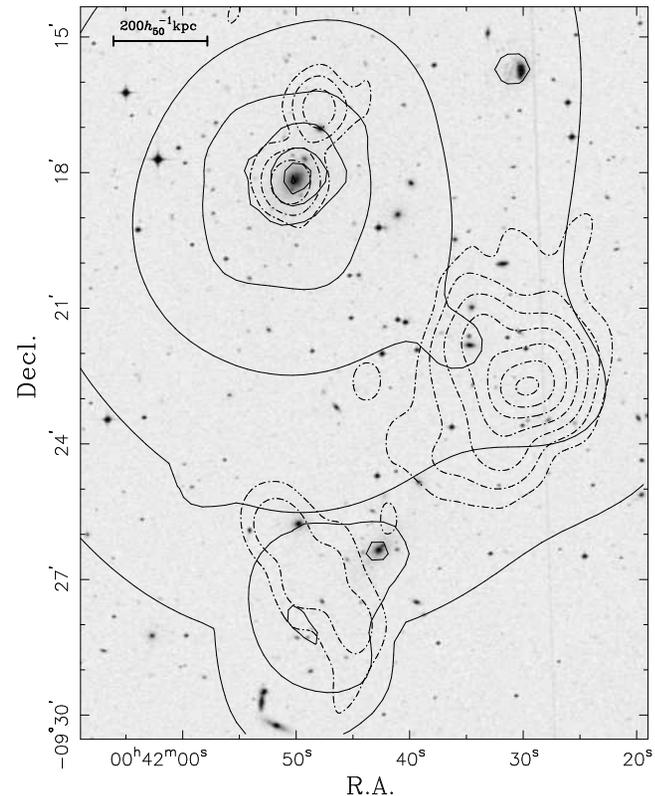,width=4in}}
\caption{\footnotesize The cluster 
Abell 85 central region at different wavelengths. The photographic
R-filter image (UK Schmidt Telescope and the Digitized Sky Survey) is shown in
grey scale. Full contour lines show the
multiscale wavelet reconstruction of the ROSAT PSPC X-ray data shown in Fig. 1.
The OSRT 326.5 MHz radio surface-brightness image is depicted using dot-dashed
contour lines. All contours are spaced logarithmically (Bagchi, Pislar, 
\& Lima Neto 1998). 
}  
\label{A85}
\end{figure}

Spatially resolving X-ray observations can also be used to separate
non-thermal and thermal X-ray emission in clusters. This technique
has been used recently in the study of the steep spectrum
radio relic source in Abell 85 \citep{bagchi99}. 
An X-ray excess relative to that expected for the cluster atmosphere
is seen with the ROSAT PSPC detector at the position
of the diffuse radio source in Abell 85 (see Fig.~\ref{A85}). 
\cite{bagchi99} subtract a model of the thermal
cluster X-ray emission in order to derive the IC contribution,
from which they derive a magnetic field of $1.0\pm0.1$  $\mu$G. 

Emission above that expected from the hot cluster atmosphere has also
been detected in the extreme ultraviolet (EUV = 0.1 to 0.4 keV) in a few
clusters \citep{berghofer01,bowyer01,bonamente01}.  It has been
suggested that this emission may also be IC in origin, corresponding
to relativistic electrons with $\gamma \sim 400$
\citep{atoyan01}. However, the emission spectrum  is steep
($\alpha \le -2$), and the EUV emitting regions are less extended
than the radio regions. Neither of these properties are consistent
with a simple extrapolation of the radio halo properties to low
frequency \citep{bowyer01}. Also, the pressure in this low $\gamma$
relativistic component would exceed that in the thermal gas by at
least a factor of three \citep{bonamente01}.

\subsection{Electron lifetimes}

An important point concerning IC and synchrotron emission from clusters
is that of particle lifetimes.  The lifetime of a relativistic electron
is limited by IC losses off the microwave background to:~ $t_{\rm IC} =
7.7\times10^9 ({{300}\over{\gamma}}) (1+z)^{-4}$ 
years \citep{Sarazin01}.\footnote{For $\gamma > 300$, IC losses
dominate (or synchrotron losses for
$\rm B > 2.3 (1+{\sl z})^2\mu$G), while for lower $\gamma$ electrons
Brehmstrahlung losses dominate in cluster
environments \citep{sarazin02}.} 
Relativistic electrons emitting  in the hard X-ray band via IC
scattering of the CMB have lifetimes of about $10^9$ years, while the
lifetimes for 1.4 GHz 
synchrotron emitting electrons 
are a factor of four or so shorter in $\mu$G fields. 
Diffusion timescales (set by streaming along the tangled magnetic
field) for cluster relativistic electrons are thought to be longer
than the Hubble time \citep{Sarazin01,colafancesco98,casse01}, 
making cluster  atmospheres efficient traps of
relativistic electrons, much like galaxy disks.
The fact that the diffusion timescales are much
longer than the energy loss timescales for $\gamma > 10^4$ electrons
requires {\sl in situ} acceleration in order to explain radio halo
sources \citep{schlikeiser87}. 

Cluster merger shock fronts are obvious sites for first order Fermi
acceleration, while subsequent turbulence may lead to second order
(stochastic) Fermi acceleration
\citep{Brunetti01b,eilek99,Ensslin98,markevitch01}. Active 
particle acceleration during cluster mergers provides a natural
explanation for the observed correlation between cluster radio halos
and substructure in cluster atmospheres \citep{govoni01}, and 
between cluster radio luminosity and
cluster atmosphere temperature, assuming that the gas temperature
increases during a merger \citep{liang00}. 
\cite{Brunetti01a} develop a two phase model in which initial 
relativistic particle injection into the ICM
occurs early in the cluster lifetime by starburst driven winds from
cluster galaxies, and/or by shocks in early sub-cluster mergers.
The second phase involves re-acceleration of the 
radiatively aged particle population via more recent cluster mergers.
Their detailed application of this model to the Coma cluster
suggests a merger has occurred within the last 10$^9$ years.

Another mechanism proposed for {\sl in situ} relativistic particle
injection is secondary electron production via the decay of
$\pi$-mesons generated in collisions between cosmic ray ions (mostly
protons) and the thermal 
ICM \citep{Dennison80,Ensslin01}.  The important point
in this case is that the energetic protons have radiative lifetimes 
orders of magnitude longer than the lower mass electrons.  
The problem with this
hypothesis is that the predicted $\gamma$-ray fluxes exceed limits set
by EGRET by a factor of 2 to 7 \citep{blasi98}.

\subsection{Reconciling IC and RM-derived fields}

The IC estimated cluster magnetic fields
are typically 0.2 to 1 $\mu$G, while those obtained
using RM observations are an order of magnitude higher.
\cite{petrosian01} has considered this discrepancy  in
detail, and finds that the different magnetic field estimates can be
reconciled in most cases
by making a few reasonable assumptions concerning the
electron energy spectrum and pitch angle distribution. 

First, an anisotropic
pitch angle distribution biased toward low angles
would clearly weaken the radio synchrotron
radiation relative to the IC X-ray emission. Such a distribution will
occur naturally due to the fact that electrons at large pitch angles
have greater synchrotron losses. A potential
problem with this argument is that pitch-angle scattering of the
relativistic electrons by Alfven waves self-induced by particles
streaming along field lines is thought to be an efficient process in
the ISM and ICM \citep{wenzel74}, such that re-isotropization of the
particle distribution will occur on a timescale short compared to
radiative timescales.  Petrosian points out that most 
derivations of magnetic fields from IC emission assume the
electrons are gyrating perpendicular to the magnetic field lines. Just
assuming an isotropic relativistic electron pitch angle distribution
raises the IC-estimated magnetic field by a factor of two or so.  

And second, the IC hard X-ray emission is from relativistic electrons
with $\gamma \sim 5000$, corresponding to radio continuum emission
at 100 MHz for $\mu$G magnetic fields.  Most surveys for cluster radio
halos have been done at 1.4 GHz \citep{Giovannini99,govoni01},
corresponding to electron Lorentz factors of $\gamma \sim 18000$.  A
steepening in the electron energy spectrum at Lorentz factors around
10$^4$ will reduce the 1.4 GHz radio luminosities, but retain the IC
hard X-ray emission. For example, Petrosian finds that a steepening in
the power-law index for the particle energy distribution from $\Gamma
= -3$ to  $\Gamma = -5$ (corresponding to $\alpha = -1$ to $-2$) at
$\gamma \sim 10^4$ raises the
IC-estimated fields to $\sim 1\mu$G.  Such a steepening of the
electron energy spectrum at $\gamma \sim 10^4$ will arise naturally 
if no relativistic particle  injection occurs over a timescale $\sim
10^8$ years (see \S 4.1).  The problem in this 
case is the fine tuning required to achieve the break in the relevant
energy range for a number of clusters. In general, a
negatively  curved (in log-space) electron energy distribution  
will inevitably lead to IC estimated fields being lower than
those estimated from 1.4 GHz radio observations, unless a correction
is made for the spectral curvature. 

Others have pointed out that magnetic substructure, or filamentation,
can lead to a significant difference between fields estimated using
the different techniques. A large relativistic electron population can
be 'hidden' from radio observations by putting them in weak field
regions \citep{feretti01,rephaeli87,rephaeli01,goldshmidt93,rudnick00}.
A simple example of this is if the relativistic particles have a
larger radial scale-length than the magnetic fields in the cluster.
In this case, most of the IC emission will come from the weak field
regions in the outer parts of the cluster, while most of the Faraday
rotation and synchrotron emission
occurs in the strong fields regions in the inner parts of the
cluster.

Another explanation for the discrepancy between IC and RM derived
magnetic fields is to assume that the hard
X-rays are not IC in origin, in which case the IC estimates become
lower limits. A number of authors have considered high energy X-ray
emission due to non-thermal Brehmstrahlung, i.e. Brehmstrahlung
radiation from a supra-thermal tail of electrons arising via
stochastic acceleration in a turbulent medium (Fermi II acceleration)
\citep{blasi00,Sarazin01,Ensslin01,dogiel00}.  The problem with
this hypothesis is the energetics: Brehmstrahlung is an inefficient
radiation mechanism, with most of the collisional energy going into
heat. Assuming an energy source is available to maintain the
supra-thermal tail, \cite{petrosian01} shows that the
collisional energy input by the supra-thermal particles would be
adequate to evaporate the cluster atmosphere on a timescale of order
10$^8$ years.

The current hard X-ray spectroscopic observations are limited to very
low spatial resolution ($\sim 1^o$), while X-ray imaging instruments
have high energy cut-offs at around 10 keV.  Likewise, sensitive,
arcminute resolution radio images for a large number of clusters are
available only at 1.4 GHz, corresponding to electrons with Lorentz
factors 3 to 4 times higher than those emitting  hard X-rays.
Both of these limitations will be overcome in the coming
years with the launch of hard X-ray imaging satellites such as
Constellation-X, and improvements in radio imaging capabilities at
300 MHz and below at the Very Large Array and the Giant Meter Wave
Radio Telescope.  

\section{Cold fronts} 

It has long been noted that in order to maintain temperature gradients
in X-ray clusters the thermal conduction must be suppressed by two
orders of magnitude relative to the classical Spitzer value
\citep{Mckee90,Fabian94,Spizter62}. If not, cooler structures on scales
$\sim 0.1$ Mpc will be evaporated by thermal conduction from the hot
surrounding medium on timescales $\sim 10^7$ years. Examples of such
cooler structures in clusters include cooling flow cluster cores and
X-ray corona surrounding large galaxies \citep{Fabian94,Vikhlinin00},
with temperature differences ranging from a factor of 2 to 5 relative to
the hot ICM. 

\cite{Cowie77} show that the conductivity
can be suppressed by almost an order of magnitude below the Spitzer
value due to the development of electrostatic
fields in cases where the Coulomb mean free path (mfp) is comparable
to the scale of thermal gradients. For large scale structure in 
cluster atmospheres this reduction is not adequate since
the $\rm mfp \sim 1.2\times10^{22} ({{T}\over{5\times10^7 K}})^2
({{n}\over{0.001 cm^{-3}}})^{-1}$ cm, or just a few kpc for a typical
cluster. 

The presence of magnetic fields will reduce the conductivity in a
thermal plasma \citep{field65,Parker79,Binney81,Chandran01}.
The simple point is that the gyro radius for thermal electrons in the
ICM is  $\rm \sim 2\times10^8 ({{B\over{1\mu G}}^{-1}}) 
({{T \over{5\times10^7 K}}})$ cm, many orders of 
magnitude below the collisional mfp. \cite{Tribble93} shows
that the presence of a cluster magnetic field will lead naturally to
the development of a multiphase ICM, with thermally isolated regions
on scales set by the magnetic structures (although
cf. \cite{rosner89}).

\begin{figure}[t!]
  \centerline{\psfig{figure=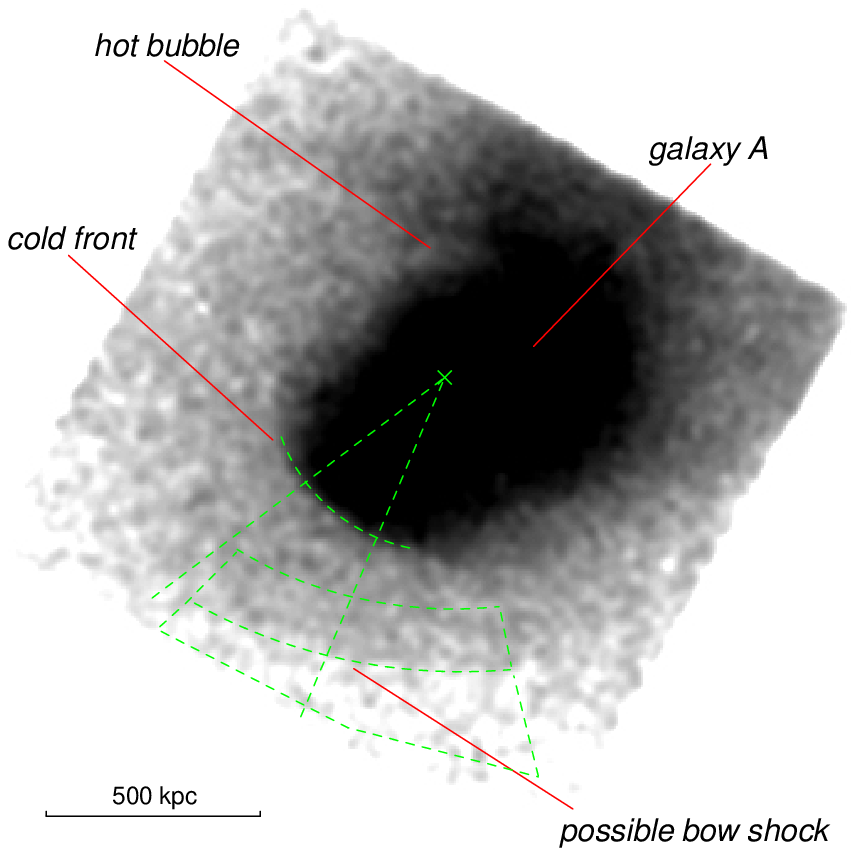,width=3.2in}}
  \centerline{\psfig{figure=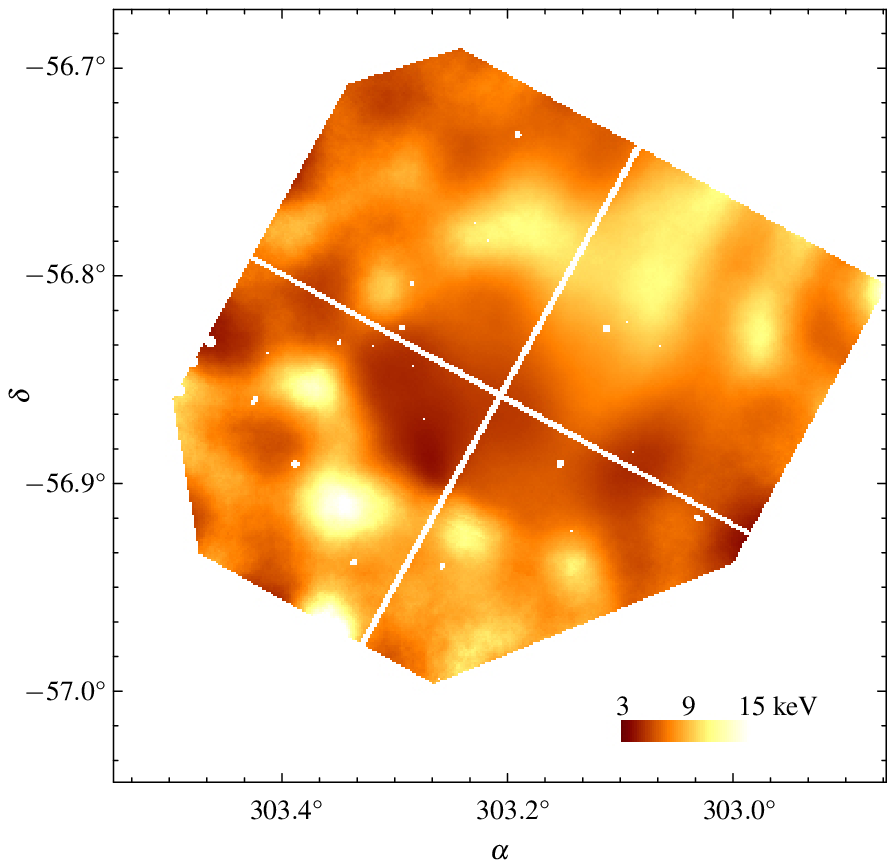,width=3.2in}}
\caption{\footnotesize (a)
    A smoothed 0.5--4 keV Chandra image of Abell 3667. The most
    prominent feature is the sharp surface brightness edge (cold front).
    The front shape is nearly circular as indicated by the arc. 
    (b) Temperature map.  The typical statistical error in this image is
    $\pm1$ keV. The cold, $\sim 4$ keV, region near the center of the map
    coincides with the inside of the cold front (Vikhlinin et al. 2001a).
}
\label{coldfront}
\end{figure}

The idea of magnetic suppression of thermal conductivity in cluster
gas has been verified with the recent discovery of `cold fronts' in
clusters of galaxies \citep{Markevitch00,
Vikhlinin00,Vikhlinin01}. These fronts
manifest themselves as sharp discontinuities in X-ray surface
brightness (Fig.~\ref{coldfront}).  They are not shocks, since the
increase in density is accompanied by a decrease in temperature such
that there is no dramatic change in 
the pressure and entropy across the front
\citep{Markevitch00,Ettori01}. These structures are interpreted as
resulting from cluster mergers, where a cooler subcluster core falls
into a hot ICM at sub- or trans-sonic velocities ($\sim 10^3$ km
s$^{-1}$). A discontinuity is formed where the internal pressure of
the core equals 
the combined ram and thermal pressure of the external medium. Gas
beyond this radius is stripped from the merging subcluster, and the
core is not penetrated by shocks due to its high internal pressure. 


The best example of a cluster cold front is that seen in Abell 3667
(see Fig. 8) \citep{Vikhlinin00,Vikhlinin01}. In this case, the
temperature discontinuity occurs
over a scale of $\sim 5$ kpc, comparable to the collisional mfp,
thereby requiring thermal isolation. Magnetic fields play a
fundamental role in allowing for such structures in two ways: (i) by
suppressing thermal conduction, and (ii) by suppressing
Kelvin-Helmholtz mixing along the contact discontinuity. 
\cite{Vikhlinin01} present a model in which the field is
tangentially sheared by fluid motions along the contact
discontinuity. They invoke magnetic tension to suppress mixing, and
show that the required magnetic pressure is between 10$\%$ and 20$\%$ of
the thermal pressure. The implied fields are between 7 and 16 $\mu$G.
They also argue that the fields cannot be much stronger than this,
since dynamically dominant fields would suppress mixing along the
entire front, which does not appear to be the case.

The existence of cold fronts provides strong evidence for cluster
magnetic fields. However, the field strengths derived correspond to
those in the tangentially sheared boundary region around the front.
Relating these back to the unperturbed cluster field probably requires
a factor of a few reduction in field strength, implying unperturbed field
strengths between 1 and 10 $\mu$G, although the exact scale factor
remains uncertain \citep{Vikhlinin01}.

\section{Magnetic support of cluster gas and the baryon crisis}

Two interesting issues have arisen in the study of cluster gas and
gravitational masses. First is the fact that total gravitating masses
derived from weak gravitational lensing are a factor of a few higher than
those derived from X-ray observations of cluster atmospheres 
assuming hydrostatic equilibrium of isothermal atmospheres
\citep{Loeb94,miralda95}.  And second is 
the `baryon crisis', in which the baryonic mass of a cluster, which is
dominated by the mass of gas in the hot cluster atmosphere,
corresponds to roughly 5$\%$ of the gravitational mass derived assuming
hydrostatic equilibrium for an isothermal cluster atmosphere. This 
baryon fraction is a factor of three to five larger than 
the baryon fraction dictated by big bang
nucleosynthesis in inflationary world models \citep{White93}.

A possible solution to both these problems is to invoke non-thermal
pressure support  for the cluster atmosphere, thereby
allowing for larger gravitating masses relative to those derived
assuming hydrostatic equilibrium.  A number of authors have
investigated the possibility of magnetic pressure support for cluster
atmospheres \citep{Loeb94,miralda95,dolag00}. The required fields are
about 50$\mu$G, which is an order of magnitude, or more,  larger than
the measured fields in most cases, except perhaps in the
inner 10's of kpc of cooling flow clusters. For most relaxed clusters
\cite{dolag00} find that magnetic pressure affects hydrostatic mass
estimates by at most 15$\%$. 

Other mechanisms for non-thermal pressure
support of cluster atmospheres involve motions of the cluster gas
other than thermal, such as turbulent or bulk motions due to a recent
cluster merger \citep{Mazzotta01,wu00}. For relaxed clusters
a number of groups have shown that the lensing and X-ray mass
estimates can be reconciled by using non-isothermal models for
the cluster atmospheres, i.e., by allowing for radial temperature 
gradients \citep{allen01b,markevitch99}. 

\section{GZK limit}

\cite{greisen66} and \cite{zatsepin} pointed out that EeV cosmic
rays lose energy due to photo-pion production through  interaction
with the CMB. These losses limit the propagation distance 
for $10^{20}$ eV particles to about 
60 Mpc. Yet no clear correlation has been
found between the arrival direction of high energy cosmic rays and
the most likely sites of origin for EeV particles, namely,  AGN
at distances less than 60 Mpc \citep{elbert95,biermann96}. 
One solution to the GZK paradox is to assume the energetic particles
are isotropized in the IGM by tangled magnetic fields, effectively
randomizing their observed arrival direction. Such isotropization
requires fields in the local super-cluster $\ge 0.3 \mu$G
\citep{farrar00,isola01}. 

\section{Synthesis}

In Table 1 we summarize the cluster magnetic field measurements.
Given the limitations of the current instrumentation, the limited
number of sources studied thus far, and the myriad physical
assumptions involved with each method, we are encouraged by the
order-of-magnitude agreement between cluster field strengths derived
from these different  methods.  Overall, the data are
consistent with cluster atmospheres containing $\sim \mu$G fields, 
with perhaps an order of magnitude scatter in field strength between
clusters, or within a given cluster.

The rotation measure observations of background radio sources, and in
particular the observations of a complete X-ray selected sample of
clusters by \cite{Clarke01} dictate that $\mu$G
magnetic fields with high areal filling factors are a standard
feature in clusters, and that the fields extend to large radii (0.5
Mpc or more).  The rotation measure observations of extended radio
galaxies embedded in clusters impose order on the fields, with
coherence scales of order 10 kpc, although larger scale coherence
in overall RM sign can be seen in some sources. Observations of inverse
Compton emission from a few clusters with radio halos provide 
evidence against much stronger, pervasive and highly tangled fields.

In most clusters the fields are not dynamically important, with
magnetic pressures one to two orders of magnitude below thermal gas
pressures. But the fields play a fundamental role in the energy budget
of the ICM through their effect on heat conduction, as is dramatically
evident in high resolution X-ray observations of cluster cold
fronts. 

If most clusters contain $\mu$G magnetic fields, then why don't most
clusters have radio halos?  The answer may be the short lifetimes of
the relativistic electrons responsible for the synchrotron radio
emission (see \S 4.1).  Without re-acceleration or injection of
relativistic electrons, a synchrotron halo emitting at 1.4 GHz will
fade in about 10$^8$ years due to synchrotron and inverse Compton
losses. This may explain the correlation between radio halos and
cluster mergers, and the anti-correlation between radio halos and
clusters with relaxed X-ray morphologies.  In this case, the fraction
of clusters with radio halos should increase with decreasing 
survey frequency.

The existence of $\mu$G-level fields in cluster atmospheres
appears well established. The challenge for observers now becomes 
one of determining the detailed properties of the fields, and how
they relate to other cluster properties. 
Are the fields filamentary, and to what extent do the
thermal and non-thermal plasma mix in cluster atmospheres?
What is the radial dependence of the field strength? 
How do the fields depend on  cluster atmospheric parameters, such
as gas temperature, metalicity, mass, substructure, 
or density profile? 
How do the fields evolve with cosmic time? 
And do the fields extend to even larger radii, perhaps 
filling the IGM? The challenge to the theorists is simpler: 
how were these field generated? This topic is considered briefly in
the next section. 

\begin{table}
{\footnotesize
\def~{\hphantom{0}}
\caption{Cluster Magnetic Fields}
\label{tab1}
\begin{tabular}{ccc}
\botrule
\botrule
Method & Strength & Model Parameters  \\
~ & $\mu$G & ~ \\
\botrule
Synchrotron Halos & 0.4 -- 1 & minimum energy, $k = \eta = 1$,\\
 & & $\nu_{\rm low}= 10$MHz, $\nu_{high}= 10$GHz \\
Faraday rotation (embedded) & 3 -- 40 & cell size = 10 kpc \\
Faraday rotation (background) & 1 -- 10 & cell size = 10 kpc \\
Inverse Compton & 0.2 -- 1 & $\alpha = -1$, $\rm \gamma_{radio} \sim
18000$,  $\rm \gamma_{xray} \sim 5000$ \\
Cold Fronts & 1 -- 10 & amplification factor $\sim$ 3 \\
GZK & $> 0.3$ & AGN = site of origin for EeV CRs \\
\botrule
\end{tabular}
}
\end{table}

\section{Field origin}

When attempting to understand the behavior of cosmic magnetic
fields, a critical characteristic to keep in mind is their
longevity. The Spitzer conductivity (Spitzer 1962) of the ICM is:
$\sigma \sim 3\times 10^{18}$ sec$^{-1}$ (for comparison, the
conductivity of liquid mercury at room temperature is 10$^{16}$
sec$^{-1}$). The timescale for magnetic diffusion in the ICM is
then: $\rm \tau_{diff} = 4 \pi \sigma ({{L}\over{c}})^2 \sim 10^{36}
({{L} \over{10 kpc}})^2$ years, where L is
the spatial scale for magnetic fluctuations. The magnetic Reynold's
number is: $\rm R_{\rm m} = {{\tau_{diff}}\over{\tau_{conv}}} \sim
10^{29} ({{L}\over{10 kpc}}) ({{V}\over{1000 km~ s^{-1}}})$,
where $\rm \tau_{conv}$ = the convective timescale $\rm = {L\over V}$, 
and  V is the bulk fluid velocity. The essentially infinite diffusion 
timescale for the fields implies that once a field is 
generated within the ICM, it will remain extant unless some
anomalous resistive process occurs {\it e.g.}, reconnection via
plasma wave generation in shocks.  

Perhaps the simplest origin for cluster magnetic fields is 
compression of an intergalactic field. Clusters have present day
overdensities $\delta \sim 10^3$. In order to get $B_{\rm ICM} >
10^{-7}$G by adiabatic compression ($B \propto \delta^{2\over 3}$)
then requires IGM fields $B_{\rm IGM} > 10^{-9} \mu$G. 

Of course, this solution merely pushes the field origin problem from
the ICM into the IGM.  An upper limit to IGM fields of $10^{-9}$G
is set by Faraday rotation measurements of high $z$ radio loud QSOs,
assuming a cell size of order 1 Mpc \citep{Kronberg96, blasi99}.  A
limit to IGM magnetic fields at the time of recombination can also be
set by considering their affect on the CMB. 
Dynamically significant magnetic fields will exert
an anisotropic pressure on the gas, which must be balanced by gravity.
Detailed studies of this phenomenon in the context of recent
measurements of the CMB anisotropies shows that the comoving IGM
fields\footnote{Comoving fields correspond to
equivalent present epoch field strengths, i.e. corrected for cosmic 
expansion assuming flux freezing.} must be less
than a few$\times 10^{-9}$ G \citep{barrow01,clarkson01,adams96}.  
A co-moving field of $10^{-9}$ G at recombination would lead to 
Faraday rotation of the polarized CMB emission by 1$^o$ at an
observing frequency of 30 GHz, a measurement that is within reach of
future instrumentation \citep{kosowsky96,grasso00}. Considerations of
primordial nucleosynthesis and the affect of magnetic fields on weak
interactions and electron densities imply upper limits to comoving
IGM fields of $10^{-7}$ G \citep{grasso95}.

The origin of IGM magnetic fields has been considered by many
authors. One class of models involves large scale field generation
prior to recombination.  An excellent review of pre-recombination
magnetic field generation is presented by \citep{grasso00}.  Early
models for pre-recombination field generation involved the
hydrodynamical (`Biermann') battery effect \citep{Biermann50}. In
general, the hydrodynamic battery involves charge separation arising
from the fact that electrons and protons have the same charge, but
very different masses. For instance, protons will have larger Coulomb
stopping distances than electrons, and be less affected by photon
drag.  \cite{harrison70} suggested that photon drag on protons
relative to electrons in vortical turbulence during the radiation era
could lead to charge separation, and hence magnetic field generation
by electric currents.  Subsequent authors have argued strongly against
vortical density perturbations just prior to recombination, since
vortical (and higher order) density perturbations decay rapidly with
the expansion of the universe \citep{rees87}. This idea has been
revisited recently in the context of vortical turbulence generated by
moving cosmic strings \citep{vachaspati91,avelino95}.  Other mechanisms
for field generation prior to recombination include battery affects
during the quark-hadron (QCD) phase transition \citep{Quashnock89},
dynamo mechanisms during the electro-weak (QED) phase transition
\citep{baym96}, and mechanisms relating to the basic physics of
inflation \citep{turner88}.

A problem with all these mechanisms is the survivability 
of the fields on relevant scales
during the radiation era. 
\cite{battaner00} argue that  magnetic and photon diffusion 
will destroy fields on comoving scales $\le$ few Mpc during this
epoch, thereby  requiring generation of the fields in the 
post-recombination universe  by normal
plasma processes during proto-galactic evolution (see also
\cite{lesch98}).

Models for post-recombination IGM magnetic field generation typically
involve ejection of the fields from normal or active galaxies
\citep{Kronberg96,Rees89}.  A simple but cogent argument in this case
is that the metalicity of the ICM is typically about 30$\%$ solar,
implying that cluster atmospheres have been polluted by outflows from
galaxies \citep{aguirre01}. A natural extension of this idea would be
to include magnetic fields in the outflows \citep{goldshmidt93}. 
It has also been suggested that IGM fields could be generated
through turbulent dynamo processes and/or shocks occurring during
structure formation 
\citep{zweibel88,kulsrud97,waxman00}, or by battery effects during the
epoch of reionization \citep{gnedin00}.

Seed magnetic fields will arise in the earliest stars via the normal
gas kinematical Biermann battery mechanism.  These
fields are amplified by the $\alpha-\Omega$ dynamo operating in stellar
convective atmospheres \citep{Parker79}, and then are ejected into the
ISM by stellar outflows and supernova explosions. The ISM fields can
then be injected into the IGM by winds from active star forming
galaxies \citep{heckman01}.  \cite{kronberg00} consider
this problem in detail and show that a population of dwarf starburst
galaxies at $z \ge 6$ could magnetize almost
50$\%$ of the universe, but that at lower redshifts the IGM volume is
too large for galaxy outflows to affect a significant fraction of the
volume.  

\cite{deyoung92} and \cite{rephaeli88} show
that galaxy outflows,
and/or gas stripping by the ICM, in present day  clusters are
insufficient to be solely responsible for cluster fields $\sim 1 \mu$G
without invoking subsequent dynamo amplification of the fields by about
an order of magnitude in the cluster atmosphere. A 
simple argument in this case is that the mean density ratio of the ICM
versus the ISM, $ \delta \sim 0.01$, such that ICM fields
would be weaker than ISM fields by $\delta^{2\over 3} \sim 0.05$,
corresponding to maximum ICM fields of 0.2 to 0.5 $\mu$G.

Fields can be ejected from Active Galactic Nuclei (AGN) by
relativistic outflows (radio jets) and Broad Absorption Line outflows
(BALs) \citep{rees68,daly90}.  The ultimate origin of the fields in
this case may be a seed field generated by a gas kinematic battery
operating in the dense accretion disk around the massive black hole,
plus subsequent amplification by an $\alpha-\Omega$ dynamo in the
rotating disk \citep{colgate00}.  Detailed consideration of this
problem \citep{furlanetto01,kronberg01} using the statistics for high
$z$ QSO populations shows that by $z \sim 3$, between 5$\%$ and 20$\%$
of the IGM may be permeated by fields with energy densities
corresponding to $\ge 10\%$ the thermal energy density of
the photo-ionized IGM at $10^4$ K, corresponding to comoving field
strengths of order 10$^{-9}$ $\mu$G.  

\cite{kronberg01} point out that powerful double radio
sources such as Cygnus A (radio luminosities $\sim
10^{45}$ erg s$^{-1}$) typically have total magnetic energies of about
10$\%$ that of the ICM as a whole. Hence, 
about ten powerful double radio sources over a cluster lifetime would
be adequate to magnetize the cluster at the $\mu$G level.

Galaxy turbulent wakes have been proposed as a means of amplifying
cluster magnetic fields \citep{jaffe80,Tribble93,ruzmaikin89}.
The problem in this case is that the energy appears to be
insufficient, with expected field strengths of at most $\sim 0.1\mu$G.
Also, the size scale of the dominant magnetic structures is 
predicted to be significantly smaller than the 5 to 10 kpc 
scale sizes observed \citep{goldshmidt93,deyoung92}.

Cluster mergers are the most energetic events in the universe since
the big bang, releasing of order 10$^{64}$ ergs in gravitational
binding energy \citep{Sarazin01}. For comparison, the total thermal
energy in the cluster atmosphere is  \\
$\rm \sim 10^{63}
({{{M_{gas}}\over{10^{14} M_\odot}}}) ({T\over{5\times 10^7 K}})$ ergs,
and the total energy contained 
in the cluster magnetic fields is $\sim 10^{60} ({B\over{1\mu
G}})^2$ ergs. Hence, only a fraction of a percent of the cluster merger
energy need be converted into magnetic fields. One possibility for
merger-generated magnetic fields is a rotational dynamo
associated with helical turbulence driven by off-center
cluster mergers.  This mechanism requires net cluster rotation --
a phenomenon that has yet to be seen in cluster galaxy velocity
fields (cf. \citep{Bregman01}).
The lack of observed rotation for  clusters suggests
low impact parameters for mergers ($\le 100$ kpc) on average
\citep{Sarazin01}, as might arise
if most mergers occur along filamentary large scale structure 
\citep{evrard01}.  The energetics of even slightly off-center cluster
mergers is adequate to generate magnetic fields at the level observed,
but the slow cluster rotation velocities ($\le$ 100 km s$^{-1}$)
imply only one or two rotations in a Hubble time \citep{colgate00},
which is insufficient for mean field generation via the
inverse cascade of the $\alpha - \Omega$ dynamo \citep{Parker79}. 

A general treatment of the  problem of magnetic field evolution
during cluster formation comes from 
numerical studies of heirarchical merging of large scale structure
including an initial intergalactic field $\sim 10^{-9}$ G
(\citep{dolag00,roettiger99}. These studies show that a combination of
adiabatic compression and non-linear amplification in shocks
during  cluster mergers may lead to ICM mean fields of order 1$\mu$G.

A related phenomenon is field amplification by (possible) cooling flows.
\cite{soker90} have considered this mechanism in
detail, and show that the amplification could be a factor of ten or
larger in the inner
10's of kpc. They predict a strong increase in RMs
with radius ($\propto r^2$), with centrally peaked radio
halos. Such an increase may explain the extreme RM values seen in
powerful radio sources at the centers of cooling flow clusters 
(see \S 3.1), although the existence of gas inflow in these systems
remains a topic of debate \citep{binney01}. 

Overall, there are a number of plausible methods for generating
cluster magnetic fields, ranging from injection of fields
into the IGM (or early ICM)
by active star forming galaxies and/or radio jets at high redshift, 
to field amplification by cluster mergers. It is likely
that a combination of these phenomena give rise to the $\mu$G fields
observed in nearby cluster atmospheres.
Tests of these mechanisms will require observations of 
(proto-) cluster atmospheres at high redshift, and
a better understanding of the general IGM field.

\section{Acknowledgements}

We thank Juan Uson and Ken Kellermann for suggesting this review
topic.  We are grateful to Stirling Colgate, Steve Cowley, Luigina
Feretti, Bill Forman, Gabriele Giovannini, Federica Govoni, Avi Loeb,
Hui Li, Vahe Petrosian, and Robert Zavala for insightful corrections
and comments on initial drafts of this manuscript. We also thank Rick
Perley, John Dreher, and Frazer Owen for fostering our initial studies
of cluster magnetic fields.  We thank G. Giovannini, T. Clarke, 
J. Bagchi, Y. Rephaeli, and A. Vikhlinin for permission to reproduce
some of the figures shown in this review.  Finally, we thank Wayne Hu
for the style files used to make this preprint.  

The National Radio Astronomy Observatory is a facility of the National
Science Foundation operated under a cooperative agreement by
Associated Universities, Inc.  This research has made use of the
NASA/IPAC Extragalactic Database (NED) which is operated by the Jet
Propulsion Laboratory, Caltech, under contract with NASA.
This research has made extensive use of NASA's Astrophysics Data System
Bibliographic Services. 

\bigskip
\bigskip

\begin{multicols}{2}

\end{multicols}

\begin{thebibliography}{}

\bibitem[{Adams et al.}, 1996]{adams96}
Adams J, Danielsson UH, Grasso D, \& Rubinstein H.\ 1996, 
Physics Letters B, 388, 253-8

\bibitem[{Aguirre et al.}, 2001]{aguirre01}
Aguirre, A, Schaye, J, Hernquist, L, Weinberg, D., Katz, N., Gardner
J. 2001, in ``Chemical Enrichment of the Intracluster and
Intergalactic Medium,'' eds. Francesca Matteucci, Roberto
Fusco-Femiano, and Max Pettini ,  (ASP: San Francisco), in press

\bibitem[{Alexander, Brown, \& Scott} (1984)]{alexander84} 
Alexander P, Brown MT, \& Scott PF, 1984, \mnras, 209, 851 

\bibitem[{Allen et al.}, 2001a]{Allen01a} 
Allen SW.\ et al.\ 2001, MNRAS, 324, 842-58

\bibitem[{Allen, Ettori, \& Fabian}, 2001b]{allen01b} 
Allen, SW, Ettori, S, \& Fabian, AC. 2001b, MNRAS, 324, 877-890

\bibitem[{Athreya et al.}, 1998]{ramana} Athreya RM, Kapahi VK, 
McCarthy PJ, \& van Breugel W.\ 1998, \aap, 329, 809 

\bibitem[{Atoyan \& V{\" o}lk}, 2000]{atoyan01}
Atoyan AM, \& V{\" o}lk HJ.\ 2000, \apj, 535, 45-52

\bibitem[{Avelino \& Shellard}, 1995]{avelino95} 
Avelino PP, \& Shellard EPS.\ 1995, \prd, 51, 5946-9

\bibitem[{Bagchi, Pislar, \& Lima Neto}, 1998]{bagchi99} 
Bagchi J, Pislar V, \& Lima Neto GB.\ 1998, \mnras, 296, L23-8

\bibitem[{Barrow, Ferreira, \& Silk}, 1997]{barrow01}
Barrow JD, Ferreira PG, \& Silk J.\ 1997, Physical Review Letters,
78, 3610-3  

\bibitem[{Battaner \& Lesch}, 2000]{battaner00}
Battaner E, \& Lesch H.\ 2000, Anales de F\'{\i}sica, in press

\bibitem[{Baym, B{\" o}deker, \& McLerran}, 1996]{baym96}
Baym G, B{\" o}deker D, \& McLerran L.\ 1996, \prd, 53, 662-7

\bibitem[{Beck et al.}, 1996]{Beck96}
Beck R, Brandenburg A, Moss D, Shukurov A, \& Sokoloff D.\ 1996, 
ARA\&A, 34, 155-206

\bibitem[{Begelman, Blandford, \& Rees}, 1984]{begelman84} 
Begelman MC, Blandford RD, \& Rees MJ.\ 1984, Reviews of Modern Physics, 
56, 255-351

\bibitem[{Bergh{\" o}fer, Bowyer, \& Korpela}, 2000]{berghofer01} 
Bergh{\" o}fer TW, Bowyer S, \& Korpela E.\ 2000, \apj, 545, 695-700

\bibitem[{Bicknell, Cameron \& Gingold}, 1990]{Bicknell90}
Bicknell GV, Cameron RA, \& Gingold RA.\ 1990, ApJ, 357, 373-87

\bibitem[{Biermann}, 1950]{Biermann50}
Biermann L.\ 1950, Zeit. Naturforshung, 5a, 65

\bibitem[{Biermann}, 1999]{biermann96} Biermann PL.\ 1999, \apss, 
264, 423-435

\bibitem[{Binney \& Cowie}, 1981]{Binney81} 
Binney J \& Cowie LL.\ 1981, \apj, 247, 464-72 

\bibitem[Binney, 2001]{binney01}
Binney J. 2001, in ``Particles and Fields in Radio Galaxies,'' eds
RA. Laing and KM. Blundell, (ASP: San Francisco), in press

\bibitem[Blasi, 2000]{blasi00} Blasi P.\ 2000, \apjl, 532, L9-12

\bibitem[{Blasi, Burles, \& Olinto}, 1999]{blasi99} 
Blasi P, Burles S, \& Olinto AV.\ 1999, \apjl, 514, L79-82

\bibitem[{Blasi \& Colafrancesco}, 1999]{blasi98} 
Blasi P \& Colafrancesco S.\ 1999, Astroparticle Physics, 12, 169-83

\bibitem[{Blumenthal \& Gould}, 1970]{blumenthal70} 
Blumenthal GR, \& Gould RJ.\ 1970, Reviews of Modern Physics, 42, 237-70

\bibitem[{Bonamente et al.}, 2001]{bonamente01} 
Bonamente M, Lieu R, Nevalainen J, \& Kaastra JS.\ 2001, \apjl in press

\bibitem[{Bowyer, Korpela, \& Bergh{\" o}fer}, 2001]{bowyer01} 
Bowyer S, Korpela E, \& Bergh{\" o}fer T.\ 2001, \apjl, 548, L135-8

\bibitem[{van Breugel, Heckman \& Miley}, 1984]{vanBreugel84}
van Breugel W, Heckman T, \& Miley G.\ 1984, ApJ, 276, 79-91

\bibitem[{van Breugel}, 2000]{vanbreugel00} 
van Breugel WJ.\ 2000, \procspie, 4005, 83-94 

\bibitem[{Bridle \& Perley}, 1984]{Bridle84} 
Bridle AH \& Perley RA.\ 1984, ARA\&A, 22, 319-58

\bibitem[{Brunetti et al.}, 2001a]{Brunetti01a} 
Brunetti G, Setti G, Feretti L, \& Giovannini G.\ 2001, MNRAS, 320, 
365-378

\bibitem[{Brunetti et al.}, 2001b]{Brunetti01b} 
Brunetti G, Setti G, Feretti L, \& Giovannini G.\ 2001, New Astronomy, 6, 1-15

\bibitem[Buote, 2001]{Buote01}
Buote D.\ 2001, \apj, 553, L15-8

\bibitem[Burbidge, 1959]{burbidge59} Burbidge GR.\ 1959, \apj, 129, 849-51

\bibitem[Burn, 1966]{burn66} Burn BJ.\ 1966, \mnras, 133, 67-83

\bibitem[{Carilli, Perley, \& Dreher}, 1988]{Carilli88} 
Carilli CL, Perley RA, \& Dreher JH.\ 1988, ApJ, 334, L73-6

\bibitem[{Carilli, Owen, \& Harris}, 1994]{carilli94} 
Carilli CL, Owen FN, \& Harris DE.\ 1994, \aj, 107, 480-93

\bibitem[{Carilli et al.}, 1997]{Carilli97}
Carilli CL, et al.\ 1997, in Extragalactic Radio Sources, eds. Fanti and
Ekers (Kluwer), p. 159

\bibitem[{Carilli et al.}, 1997]{carilli97} Carilli, CL, 
Roettgering,HJA, van Ojik,R, Miley, GK, \& van Breugel, 
WJM.\ 1997, \apjs, 109, 1 

\bibitem[{Carilli, Perley \& Dreher}, 1988]{Carilli98a}
Carilli CL, Perley RA, \& Dreher JW.\ 1988, ApJ, 334, L73-6

\bibitem[{Carilli et al.}, 1998]{Carilli98b}
Carilli CL, Harris DE, Pentericci L, Rottergering HJA, Miley GK,
Bremer MN.\ 1998, ApJ, 494, L143-6

\bibitem[{Carilli et al.}, 2001]{carilli01}
Carilli CL. et al. 2001, in {\sl Starburst Galaxies Near and Far}, (Springer:
Berlin), in press

\bibitem[{Casse, Lemoine \& Pelletier}, 2001]{casse01}
Casse F, Lemoine M, \& Pelletier G.\ 2001,
Phys. Rev. D, in press 

\bibitem[{Cavaliere \& Fusco-Femiano}, 1976]{Cavaliere76}
Cavaliere A, \& Fusco-Femiano R.\ 1976, \aap, 49, 137-44

\bibitem[{Chambers, Miley, \& van Breugel}, 1990]{chambers89} 
Chambers KC, Miley GK, \& van Breugel WJM.\ 1990, \apj, 363, 21-39 

\bibitem[{Chandran, Cowley, \& Albright}, 1999]{Chandran01} 
Chandran BDG, Cowley SC, \& Albright B.\ 1999, in {\it Diffuse Thermal 
and Relativistic Plasma in Galaxy Clusters},
eds.  H Bohringer, L Feretti, P Schuecker,  MPE report 271, pp. 242-6

\bibitem[{Clarke, Kronberg, \& B{\" o}hringer}, 2001]{Clarke01} 
Clarke TE, Kronberg PP, \& B{\" o}hringer H.\ 2001, ApJ, 547, L111-4

\bibitem[{Clarkson \& Coley}, 2001]{clarkson01} Clarkson CA. \& 
Coley AA. 2001, Classical Quantum Gravity, 18, 1305-1310

\bibitem[{Colafrancesco \& Blasi}, 1998]{colafancesco98} 
Colafrancesco S, \& Blasi P.\ 1998, Astroparticle Physics, 9, 227-46

\bibitem[{Colgate \& Li}, 2000]{colgate00} 
Colgate SA, \& Li H.\ 2000,
in {\sl Highly Energetic Physical Processes and Mechanisms for
Emission from Astrophysical Plasmas, Proceedings of IAU Symposium
\#195}, (ASP: San Francisco), 255-65

\bibitem[{Condon et al.}, 1998]{Condon98}
Condon JJ, Cotton WD, Greisen EW, Yin QF, Perley RA, Taylor GB, \& 
Broderick JJ.\ 1998, AJ, 115, 1693-1716

\bibitem[{Cowie \& McKee}, 1977]{Cowie77}  
Cowie LL, \& McKee CF.\ 1977, \apj, 211, 135-146

\bibitem[{Daly \& Loeb}, 1990]{daly90}
Daly RA, \& Loeb A.\ 1990, \apj, 364, 451-5  

\bibitem[{Dennison}, 1980]{Dennison80} Dennison B.\ 1980, ApJ, 239, L93-6

\bibitem[{De Young}, 1992]{deyoung92}
De Young DS.\ 1992 \apj, 386, 464-72

\bibitem[{Dogiel}, 2000]{dogiel00}
Dogiel VA.\ 2000, A\&A, 357, 66-74

\bibitem[{Dolag \& Schindler}, 2000]{dolag00}
Dolag K, \& Schindler S.\ 2000, A\&A, 364, 491-6 

\bibitem[{Dolag et al.}, 2001]{Dolag01}
Dolag K, Schindler S, Govoni F, \& Feretti L. 2001, A\&A, in press

\bibitem[{Dreher, Carilli \& Perley}, 1987]{Dreher87}
Dreher JW, Carilli CL, \& Perley RA.\ 1987, ApJ, 315, 611-25

\bibitem[{Dupke \& Bregman}, 2001]{Bregman01} 
Dupke RA, \& Bregman JN.\ 2001, \apj, 547, 705-13

\bibitem[{Ebeling et al.}, 1996]{Ebeling96}
Ebeling H, Voges W, B\"ohringer H, Edge AC, Huchra JP, \& Briel UG.\ 1996
MNRAS, 281, 799

\bibitem[{Eilek}, 1999]{eilek99}
Eilek JA.\ 1999, in {\sl Diffuse thermal and relativistic plasma in
galaxy clusters}, eds.  H Bohringer, L Feretti, P Schuecker, 
MPE Report 271, pp. 71-6

\bibitem[{Eilek \& Owen}, 2001]{Eilek01}
Eilek JA, \& Owen FN.\ 2001, ApJ, in press

\bibitem[{Elbert \& Sommers}, 1995]{elbert95}
Elbert JW, \& Sommers P.\ 1995, \apj, 441, 151-61

\bibitem[{Ensslin et al.}, 1998]{Ensslin98} 
Ensslin TA, Biermann PL, Klein U, \& Kohle S.\ 1998, A\&A, 
332, 395-409

\bibitem[{Ensslin \& Gopal-Krishna}, 2001]{Ensslin01}
Ensslin TA, \& Gopal-Krishna 2001, A\&A, 366, 26-34

\bibitem[{Ettori \& Fabian}, 2000]{Ettori01}
Ettori S, \& Fabian AC. 2000, \mnras, 317, L57-9

\bibitem[{Evrard \& Gioia}, 2001]{evrard01}
Evrard AE \& Gioia IM.\ 2001 in {\sl Merging processes in
clusters of galaxies}, ed. L. Feretti, I.M. Gioia, \& G. Giovannini
(Dordrecht: Kluwer) in press.

\bibitem[{Fabbiano et al.}, 1979]{Fabbiano79} 
Fabbiano G, Schwartz DA, Schwarz J, Doxsey RE, \& Johnston M.\ 1979, 
ApJ, 230, L67-71

\bibitem[{Fabian, Nulsen, \& Canizares}, 1991]{Fabian91} 
Fabian AC, Nulsen PEJ, \& Canizares CR.\ 1991, A\&A Reviews, 2, 191-226 

\bibitem[{Fabian}, 1994]{Fabian94}
Fabian AC.\ 1994, ARA\&A, 32, 277

\bibitem[{Fanaroff \& Riley}, 1974]{Fanaroff74}
Fanaroff BL, \& Riley JM.\ 1974, \mnras, 167, L31-5

\bibitem[Farrar, 2000]{farrar00}
Farrar GR, \& Piran T.\ 2000, Phys. Rev.Lett. 84, 3527-30

\bibitem[{Feigelson et al.}, 1995]{feigelson95}
Feigelson, ED, Laurent-Muehleisen, SA, Kollgaard, RI, \& Fomalont,
EB 1995, \apjl, 449, L149-152

\bibitem[{Field}, 1965]{field65}
Field GB.\ 1965, \apj, 142, 531-67

\bibitem[Felten, 1996]{Felten96}
Felten JB.\ 1996, in ``Clusters, Lensing and the
Future of the Universe'' eds. V. Trimble \& A. Reisenegger, 
ASP Conf. Series, Vol.~88, pp. 271-3  

\bibitem[{Feretti et al.}, 1995]{Feretti95} 
Feretti L, Dallacasa D, Giovannini G, \& Tagliani A.\ 1995, A\&A, 302, 680-90 

\bibitem[{Feretti et al.}, 1998]{Feretti98}
Feretti L, \& Giovannini G.\ 1998 in {\it A new view of an
old Cluster: untangling Coma Berenices}, Eds. A. Mazure, F. Casoli, F.
Durret, \& D. Gerbal, Word Scientific Publishing Co Pte Ltd, p. 123

\bibitem[{Feretti}, 1999]{Feretti99b}
Feretti L.\ 1999, in {\sl Diffuse thermal and relativistic plasma in
galaxy clusters}, eds.  H Bohringer, L Feretti, P Schuecker, MPE
report 271, pp. 3-8

\bibitem[{Feretti et al.}, 1999]{Feretti99} 
Feretti L, Dallacasa D, Govoni F, Giovannini G, Taylor GB, \& 
Klein U.\ 1999, A\&A, 344, 472-82 

\bibitem[{Feretti et al.}, 2001]{feretti01} Feretti L. Fusco-Femiano R. 
Giovannini G. \& Govoni F. 2001, A\&A, 373, 106-12

\bibitem[{Forman et al.}, 1972]{Forman72} Forman, W, Kellogg, E, 
Gursky, H, Tananbaum, H, \& Giacconi, R. 1972, \apj, 178, 309 

\bibitem[{Furlanetto \& Loeb}, 2001]{furlanetto01} 
Furlanetto S, \& Loeb A.\ 2001,  submitted to \apj

\bibitem[{Fusco-Femiano et al.}, 2000]{fusco00}
Fusco-Femiano R, et al.\ 2000, \apjl, 534, L7-10

\bibitem[{Fusco-Femiano et al.}, 2001]{fusco01}
Fusco-Femiano R, et al.\ 2001, \apjl, 552, L97-100

\bibitem[{Garrington et al.}, 1988]{Garrington88}
Garrington ST, Leahy JP, Conway RG, \& Laing RA.\ 1988, Nature, 331, 147-9

\bibitem[{Ge \& Owen}, 1993]{Ge93} 
Ge JP, \& Owen FN.\ 1993, AJ, 105, 778-87

\bibitem[{Ge \& Owen}, 1994]{Ge94}
Ge JP, \& Owen, FN.\ 1994, AJ, 108, 1523-33

\bibitem[{Giovannini et al.}, 1993]{Giovannini93} 
Giovannini G, Feretti L, Venturi T, Kim K-T, \& Kronberg PP.\ 1993, ApJ, 
406, 399-406

\bibitem[{Giovannini, Tordi \& Feretti}, 1999]{Giovannini99} Giovannini G, 
Tordi M, \& Feretti L.\ 1999, New Astronomy, 4, 141-55

\bibitem[{Giovannini \& Feretti}, 2000]{Giovannini00} Giovannini G, 
\& Feretti L.\ 2000, New Astronomy, 5, 335-47

\bibitem[{Gnedin et al.}, 2000]{gnedin00}
Gnedin NY, Ferrara A, Zweibel EG.\ 2000, \apj, 539, 505-16

\bibitem[{Goldshmidt \& Rephaeli}, 1993]{goldshmidt93}
Goldshmidt O, \& Rephaeli Y.\ 1993, \apj, 411, 518-28

\bibitem[{Govoni et al.}, 2001a]{govoni01} 
Govoni F, Feretti L, Giovannini G, Bohringer H, Reiprich TH, \& 
Murgia M. 2001a, A\&A., in press 

\bibitem[{Govoni et al.}, 2001b]{Govoni01}
Govoni F, Taylor GB, Dallacasa D, Feretti L, \& Giovannini G.\ 2001b, A\&A,
submitted

\bibitem[{Grasso and Rubinstein}, 1995]{grasso95}
Grasso D \& Rubinstein HR.\ 1995, Nuc.Phys. B, 43, 303-7

\bibitem[{Grasso and Rubinstein}, 2001]{grasso00}
Grasso D \& Rubinstein HR. 2001, Phys. Rep., 348, 163-266

\bibitem[Greisen, 1966]{greisen66}
Greisen K.\ 1966, Phys. Rev.Lett., 16, 748-58

\bibitem[Harrison, 1970]{harrison70}
Harrison ER.\ 1970, \mnras, 147, 279


\bibitem[Hanisch, 1982]{Hanisch82}
Hanisch RJ.\ 1982, A\&A, 116, 137-46

\bibitem[{Harris \& Miley}, 1978]{Harris78}
Harris DE, \& Miley G.\ 1978, A\&AS, 34, 117-28

\bibitem[{Harris \& Grindlay}, 1979]{harris79} 
Harris DE, \& Grindlay JE.\ 1979, \mnras, 188, 25-37

\bibitem[{Harris et al.}, 1993]{Harris93} 
Harris DE, Stern CP, Willis AG, \& Dewdney PE.\ 1993, AJ, 105, 769-77

\bibitem[Heckman, 2001]{heckman01} Heckman TM. 2001, 
in {\it Extragalactic Gas at Low Redshift}, eds. J. Mulchaey and 
J. Stocke, (ASP: San Francisco)

\bibitem[{Hennessy, Owen \& Eilek}, 1989]{Hennessy89}
Hennessy GS, Owen FN, \& Eilek JA.\ 1989, ApJ, 347, 144

\bibitem[{Isola, Lemoine, \& Sigl}, 2001]{isola01}
Isola C, Lemoine M, \& Sigl G.\ 2001, astro-ph/0104289 

\bibitem[Jaffe, 1977]{Jaffe77} Jaffe, W.\ 1977, IAU Symp.~74: 
Radio Astronomy and Cosmology, 74, 305 


\bibitem[Jaffe, 1980]{jaffe80} Jaffe WJ.\ 1980, \apj, 241, 925-7 

\bibitem[Jaffe, 1992]{jaffe92} Jaffe WJ.\ 1992, Clusters and 
Superclusters of Galaxies, 109 

\bibitem[{Jones et al.}, 1979]{Jones79} 
Jones C, Mandel E, Schwarz J, Forman W, Murray SS, \& Harnden FR.\ 1979,
  \apjl, 234, L21-4

\bibitem[{Kempner \& Sarazin}, 2001]{Kempner01} 
Kempner JC, \& Sarazin CL.\ 2001, ApJ, 548, 639-51

\bibitem[{Kim et al.}, 1990]{Kim90}
Kim KT, Kronberg PP, Dewdney PE, \& Landecker TL.\ 1990, 
ApJ, 355, 29-37

\bibitem[{Kim, Kronberg, \& Tribble}, 1991]{Kim91} 
Kim KT, Kronberg PP, \& Tribble PC.\ 1991, ApJ, 379, 80-8

\bibitem[{Kronberg}, 1996]{Kronberg96}
Kronberg PP.\ 1996, Space Science Reviews, 75, 387-99

\bibitem[{Kronberg, Lesch, \& Hopp}, 1999]{kronberg00} 
Kronberg PP, Lesch H, \& Hopp U.\ 1999, \apj, 511, 56-64

\bibitem[{Kronberg et al.}, 2001]{kronberg01}
Kronberg PP, Dufton QW, Li H, \& Colgate SA.\ 2001, \apj, in press

\bibitem[{Kosowsky \& Loeb}, 1996]{kosowsky96} 
Kosowsky A. \& Loeb A.\ 1996, \apj, 469, 1-6

\bibitem[{Kulsrud}, 1999]{Kulsrud99}
Kulsrud RM.\ 1999, ARA\&A, 37-64

\bibitem[{Kulsrud et al.}, 1997]{kulsrud97}
Kulsrud RM, Cen R, Ostriker JP, \& Ryu D.\ 1997, \apj, 480, 481-91

\bibitem[Large, 1959]{Large59}
Large MI.\ 1959, Nature, 183, 1663-4

\bibitem[{Lesch \& Birk}, 1998]{lesch98} 
Lesch H, \& Birk GT. 1998, Physics of Plasmas, 5, 2773-6

\bibitem[{Liang et al.}, 2000]{liang00} 
Liang H, Hunstead RW, Birkinshaw M, \& Andreani P. 2000, \apj, 544, 686-701

\bibitem[{Lieu et al.}, 1999]{lieu99} 
Lieu R, Bonamente M, Mittaz JPD, Durret F, Dos Santos S, \& Kaastra
JS.\ 1999, \apjl, 527, L77-80

\bibitem[{Loeb \& Mao}, 1994]{Loeb94}
Loeb A, \& Mao S.\ 1994, \apjl, 435, L109-12

\bibitem[{Lonsdale, Barthel, \& Miley}, 1993]{lonsdale} Lonsdale, 
CJ, Barthel, PD, \& Miley, GK. 1993, \apjs, 87, 63 

\bibitem[{Mathews \& Brighenti}, 1997]{Mathews97}
Mathews WG, \& Brighenti F.\ 1997, ApJ, 488, 595-605

\bibitem[{Markevitch \& Vikhlinin}, 2001]{markevitch01}
Markevitch M, \&  Vikhlinin, A 2000, \apj, in press

\bibitem[{Markevitch et al.}, 2000]{Markevitch00}
Markevitch M. et al. 2000, \apj, 541, 542-9

\bibitem[{Markevitch et al.}, 1999]{markevitch99}
Markevitch M, Vikhlinin, A, Forman, WR, \& Sarazin, CL. 1999, 
\apj, 527, 545-553

\bibitem[{Markovic \& Eilek}, 2001]{Markovic01}
Markovic T, \& Eilek JA.\ 2001, in preparation

\bibitem[{Mazzotta et al.}, 2001]{Mazzotta01}
Mazzotta P, Markevitch M, Vikhlinin A, Forman WR, David
LP, VanSpeybroeck L.\ 2001, \apj, 555, 205-14

\bibitem[{McKee \& Begelman}, 1990]{Mckee90}
McKee CF, \& Begelman MC.\ 1990, \apj, 358, 392-8

\bibitem[Meyer, 1969]{meyer69} 
Meyer P.\ 1969, \araa, 7, 1-38 

\bibitem[Miley, 1980]{miley80}
Miley G.\ 1980, \araa, 165-218

\bibitem[{Miralda-Escude \& Babul}, 1995]{miralda95}
Miralda-Escude J, \& Babul A.\ 1995, \apj, 449, 18-27

\bibitem[Mitton, 1971]{mitton71}
Mitton S.\ 1971, MNRAS, 153, 133-43

\bibitem[{Nodland \& Ralston}, 1997]{Nodland97}
Nodland B, \& Ralston JP.\ 1997, Phys. Rev. Lett, 78, 3043-6

\bibitem[{Owen, Morrison \& Vogues}, 1999]{owen99}
Owen F, Morrison G, \& Vogues W.\ 1999, in {\sl Diffuse thermal and relativistic plasma in
galaxy clusters}, eds.  H Bohringer, L Feretti, P Schuecker, MPE
Report 271, pp. 9-11

\bibitem[Pacholczyk, 1970]{Pacholczyk70}
Pacholczyk AG.\ 1970, {\it Radio Astrophysics}, (Freeman:San
Francisco)

\bibitem[Parker, 1979]{Parker79}
Parker EN.\ 1979, {\it Cosmical Magnetic Fields, Their Origin and
Their Activity}, (Clarendon Press: Oxford)

\bibitem[Peebles, 1993]{Peebles93}
Peebles PJE.\ 1993, {\sl Principles of Physical Cosmology} 
(Princeton Univ. Press)

\bibitem[{Pentericci et al.}, 2000]{pentericci01} 
Pentericci L, Van Reeven W, Carilli CL, R{\" o}ttgering HJA, \& Miley GK.\ 
2000, \aaps, 145, 121-59

\bibitem[{Perley \& Taylor}, 1991]{Perley91} 
Perley RA, \& Taylor GB.\ 1991, AJ, 101, 1623-31

\bibitem[Petrosian, 2001]{petrosian01} Petrosian V. 2001, \apj, in
press 

\bibitem[{Quashnock, Loeb, \& Spergel}, 1989]{Quashnock89} 
Quashnock JM, Loeb A. \& Spergel DN.\ 1989, \apjl, 344, L49-51

\bibitem[{Rees \& Setti}, 1968]{rees68}
Rees MJ, \& Setti G.\ 1968, Nature, 219, 127-31

\bibitem[Rees, 1987]{rees87} Rees MJ.\ 1987, \qjras, 28, 197-206

\bibitem[Rees, 1989]{Rees89} 
Rees MJ.\ 1989, in {\it Problems in Theoretical Physics and Astrophysics}
415-21

\bibitem[{Rephaeli, Gruber, \& Rothschild}, 1987]{rephaeli87} 
Rephaeli Y. Gruber DE. \& Rothschild RE.\ 1987, \apj, 320, 139-144

\bibitem[Rephaeli, 1988]{rephaeli88}
Rephaeli Y.\ 1988, Comm. Mod Phys., vol. 12, part C, 265-79

\bibitem[{Rephaeli}, 1995]{rephaeli95} Rephaeli Y.\ 1995, \araa, 
33, 541-580

\bibitem[{Rephaeli, Gruber, \& Blanco}, 1999]{rephaeli99}
Rephaeli Y, Gruber D, \& Blanco P.\ 1999, \apjl, 511, L21-4

\bibitem[Rephaeli, 2001]{rephaeli01}
Rephaeli Y.\ 2001, in `High Energy Gamma-Ray Astronomy'

\bibitem[{Roettiger, Burns \& Stone}, 1998]{Roettiger98}
Roettiger K, Burns JO, \& Stone JM.\ 1998, in {\sl 9th Texas
Symposium on Relativistic Astrophysics and Cosmology} eds. J Paul,
T Montmerle, \& E Aubourg

\bibitem[{Roettiger, Stone, \& Burns}, 1999]{roettiger99} 
Roettiger K, Stone JM, \& Burns JO.\ 1999, \apj, 518, 594-602

\bibitem[{Rosner \& Tucker}, 1989]{rosner89}
Rosner R, \& Tucker WH.\ 1989, \apj, 338, 761-9

\bibitem[Rudnick, 2000]{rudnick00} 
Rudnick, L.\ 2000, in `Cluster mergers and their connection to radio 
sources,' 24th meeting of the IAU, JD10, E22 

\bibitem[{Ruzmaikin, Sokolov, \& Shukurov}, 1989]{ruzmaikin89}
Ruzmaikin A, Sokolov D, \& Shukurov A.\ 1989, \mnras, 241, 1-14 

\bibitem[{Ruzmaikin,  Shukurov, \& Sokolov},  1987]{ruzmaikin87}
Ruzmaikin A, Shukurov A,  \& Sokolov D.\ 1987, Magnetic Fields of
Galaxies, (Kluwer: Dordrecht)

\bibitem[Sarazin, 1988]{Sarazin88}
Sarazin CL.\ 1988, {\sl X-ray Emission from Clusters of Galaxies}
(CUP)

\bibitem[Sarazin, 2001a]{Sarazin01} Sarazin CL.\ 2001, in
{\sl Galaxy Clusters and the High Redshift Universe Observed 
in X-rays}, eds D Neumann, F Durret, \& J Tran Thanh Van

\bibitem[Sarazin, 2001b]{sarazin02} Sarazin CL.\ 2001, in
Merging Processes in Clusters of Galaxies, eds. L Feretti, IM Gioia,
and G Giovannini (Dordrecht: Kluwer)

\bibitem[{Schlikeiser, Sievers, \& Thiemann}, 1987]{schlikeiser87}
Schlikeiser R, Sievers A, \& Thiemanns H. 1987, \aap, 182, 21-35

\bibitem[{Simard-Normandin, Kronberg \& Button}, 1981]{Simard81}
Simard-Normandin M, Kronberg PP, \& Button S.\ 1981, ApJS, 45, 97-111

\bibitem[{Smith \etal}, 1974]{Smith74}
Smith E, Jones DE, Coleman PJ, Colburn DS, Dyal P, 
Sonett PCP, \& Frandsen AMR.\ 1974, {\it J. Geophys. Res.}, 79, 3501-11

\bibitem[{Soker \& Sarazin}, 1990]{soker90} 
Soker N, \& Sarazin CL.\ 1990, \apj, 348, 73-84

\bibitem[Soward, 1983]{Soward83}
Soward AM.\ 1983, {\it Stellar and Planetary Magnetism}, (Gordon
and Breach: New York)

\bibitem[Spitzer, 1962]{Spizter62}
Spitzer L.\ 1962, The Physics of Fully Ionized Gases, (Interscience:
New York) 

\bibitem[{Spitzer}, 1978]{spitzer78}
Spitzer L.\ 1978, Physical Processes in the Interstellar Medium,
(Wiley: New York)

\bibitem[{Stern \& Ness}, 1982]{stern82} 
Stern DP, \& Ness NF.\ 1982, \araa, 20, 139-61 

\bibitem[{Taylor, Inoue, \& Tabara}, 1992]{Taylor92} 
Taylor GB, Inoue M, \& Tabara H.\ 1992, A\&A, 264, 421-7

\bibitem[{Taylor \& Perley}, 1993]{Taylor93} 
Taylor GB, \& Perley RA.\ 1993, ApJ, 416, 554-62

\bibitem[{Taylor, Barton \& Ge}, 1994]{Taylor94}
Taylor GB, Barton EJ, \& Ge JP.\ 1994, AJ, 107, 1942-52

\bibitem[{Taylor}, 1998]{Taylor98}
Taylor GB.\ 1998, ApJ, 506, 637-46

\bibitem[{Taylor et al.}, 2001a]{Taylor01a}
Taylor GB, Govoni F, Allen SW, \& Fabian AC.\ 2001a, MNRAS, in press

\bibitem[{Taylor et al.}, 2001b]{Taylor01b}
Taylor GB, Allen SW, \& Fabian AC.\ 2001b, MNRAS, submitted

\bibitem[{Tribble}, 1993]{Tribble93}
Tribble PC.\ 1993, \mnras,  263, 31-36

\bibitem[{Turner \& Widrow}, 1988]{turner88}
Turner MS, \& Widrow LM.\ 1988, \prd, 37, 2743-54

\bibitem[{Vachaspati \& Vilenkin}, 1991]{vachaspati91}
Vachaspati T, \& Vilenkin A.\ 1991, \prd, 43, 3846-55

\bibitem[{Vallee, MacLeod, \& Broten}, 1986]{Vallee86} 
Vallee JP, MacLeod JM, \& Broten NW.\ 1986, A\&A, 156, 386-90

\bibitem[{Vallee, MacLeod, \& Broten}, 1987]{Vallee87} 
Vallee JP, MacLeod JM, \& Broten NW.\ 1987, ApL, 25, L181-6

\bibitem[{Vikhlinin et al.}, 2001a]{Vikhlinin01}
Vikhlinin A, Markevitch M \& Murray SS. 2001a, \apjl, 549, L47-50

\bibitem[{Vikhlinin et al.}, 2001b]{Vikhlinin00}
Vikhlinin A, Markevitch M, Forman W \& Jones C.\ 2001b,
\apjl, 555, L87-90

\bibitem[{Wardle, Perley \& Cohen}, 1997]{Wardle97}
Wardle J, Perley RA, \& Cohen M.\ 1997, Phys. Rev. Lett, 79, 1801-4

\bibitem[{Warwick}, 1963]{Warwick63}
Warwick JA.\ 1963, ApJ, 137, 41-60

\bibitem[{Waxman \& Loeb}, 2000]{waxman00}
Waxman, Eli \& Loeb, A. 2000, \apjl, 545, 11-14

\bibitem[{Wentzel}, 1974]{wenzel74} 
Wentzel DG.\ 1974, \araa, 12, 71-96

\bibitem[{White et al.}, 1993]{White93}
White SDM, Navarro JF, Evrard AE, \& Frenk, CS.\ 1993, 
\nat, 366, 429-31 

\bibitem[{Willson}, 1970]{Willson70}
Willson MAG.\ 1970, MNRAS, 151, 1-44

\bibitem[{Wu}, 2000]{wu00}
Wu Xiang-Ping 2000, MNRAS, 316, 299-306

\bibitem[{Zatsepin \& Kuzmin}, 1966]{zatsepin}
Zatsepin, GT, \& Kuzmin, VA. 1966, Phys.-JETP Lett. 4, 78-83

\bibitem[{Zweibel}, 1988]{zweibel88} 
Zweibel EG.\ 1988, \apjl, 329, L1-4 

\end{thebibliography}
\end{document}